\documentclass[12pt,preprint,showpacs,preprintnumbers]{revtex4}
\usepackage{amsfonts}
\usepackage{amsmath}
\usepackage{amssymb}
\usepackage{hyperref}
\usepackage{color}
\usepackage{graphicx}
\usepackage{amssymb}
\usepackage{amsmath}
\usepackage{graphicx}
\usepackage{dcolumn}
\usepackage{bm}
\usepackage{epsfig}
\usepackage[T1]{fontenc}
\usepackage{ae,aecompl}

\begin{document}

\title{Polar coordinate lattice Boltzmann modeling of compressible flows}
\author{Chuandong Lin$^{1}$,
Aiguo Xu$^{2,3}$\footnote{Corresponding author: Xu\_Aiguo@iapcm.ac.cn},
Guangcai Zhang$^{2}$,
Yingjun Li$^{1}$\footnote{Corresponding author: lyj@aphy.iphy.ac.cn},
Sauro Succi$^{4}$}
\affiliation{$^1$ State Key Laboratory for GeoMechanics and Deep Underground Engineering,
China University of Mining and Technology, Beijing 100083, P.R.China\\
$^2$ National Key Laboratory of Computational Physics,
Institute of Applied Physics and Computational Mathematics, P. O. Box 8009-26, Beijing 100088, P.R.China \\
$^3$ Center for Applied Physics and Technology, MOE Key Center for High Energy Density Physics Simulations, College of Engineering, Peking University, Beijing 100871, China \\
$^4$ Istituto Applicazioni Calcolo-CNR - Viale del Policlinico 137, 00161, Roma, Italy, EU
 }
\date{\today}

\begin{abstract}
We present a Polar Coordinate Lattice Boltzmann kinetic model for compressible flows. A method to recover the continuum distribution function from the discrete distribution function is indicated. Within the model, a hybrid scheme being similar to, but different from, the operator-splitting is proposed. The temporal evolution is calculated analytically and the convection term is solved via a Modified Warming-Beam (MWB) scheme. Within the MWB scheme a suitable switch function is introduced. The current model works not only for subsonic flows but also for supersonic flows. It is validated and verified via the following well-known benchmark tests: (i) the rotational flow, (ii) the stable shock tube problem, (iii) the Richtmyer-Meshkov (RM) instability, (iv) the Kelvin-Helmholtz instability. As an original application, we studied the non-equilibrium characteristics of the system around three kinds of interfaces, the shock wave, the rarefaction wave and the material interface, for two specific cases. In one of the two cases, the material interface is initially perturbed and consequently the RM instability occurs. It is found that, the macroscopic effects due to deviating from thermodynamic equilibrium around the material interface differ significantly from those around the mechanical interfaces. The initial perturbation at the material interface enhances the coupling of molecular motions in different degrees of freedom. The amplitude of deviation from thermodynamic equilibrium around the shock wave is much higher than those around the rarefaction wave and material interface. By comparing each component of the high-order moments and its value in equilibrium, we can draw qualitatively the main behavior of the actual distribution function. These results deepen our understanding of the mechanical and material interfaces from a more fundamental level, which is indicative for constructing macroscopic models and other kinds of kinetic models.
\end{abstract}

\maketitle

\section{\protect\bigskip Introduction}

During recent decades the lattice Boltzmann (LB) modeling and simulation have achieved great success in various complex flows \cite{Succi-Book}. However, most of these studies were focused on nearly incompressible flow, while with increasing the Mach number, the compressibility of flows has to be taken into account. Such high speed compressible flows are ubiquitous in aerophysics, astrophysics, explosion physics, medical physics and others. Given the great importance of shock waves in many fields of physics and engineering, constructing LB models for high speed compressible flows has attracted considerable interest since the early days of LB research \cite{Succi-Book}.

In 1992 Alexander et al \cite{Alexander1992} formulated a compressible LB model for flows at high Mach number via introducing a flexible sound speed. This model works only for nearly isothermal compressible systems. In 1999 Yan et al \cite{Yan1999} proposed a LB scheme for compressible Euler equations. In the years of 1998 and 2003 Sun and his coworker \cite{Sun1998,Sun2003} presented an adaptive LB scheme for the two- and three-dimensional systems, respectively. In this model the particle velocities vary with the Mach number and internal energy, so that the particle velocities are no longer constrained to fixed values. All of those models belong to the standard LB framework. However, due mainly to numerical instability problems, applications of LB methods to compressible flows remain scanty to date.

Besides the standard LB framework, the other way to formulate LB for high speed flows is to use the Finite-Difference (FD) scheme to calculate the temporal and spatial derivatives of the distribution function. In 1997 Cao et al \cite{Cao1997} proposed to use the FD scheme to improve the numerical stability and apply nonuniform grids in the LB method. In the past decade, Tsutahara, Watari and Kataoka \cite{Kataoka2004a,Kataoka2004b,Watari2003,Watari2004,Watari2007} proposed several nice FDLB models for the Euler and Navier-Stokes equations, where the discretizations in the space and in particle velocity are separated. In 2005 Xu \cite{Xu2005PRE,Xu2005EPL} extended the idea to handle binary fluids. However, similar to the case of standard LB models, these FDLB schemes only work for subsonic flows. Physical simulation and practical application are the goals of LB method \cite{XGL1,XGL2,XGL3,LB2MPhase2011PRE,LB2MPhase2012EPL,LB2MPhase2012FrontPhys,
XuPan2007,XuGan2008CTP,XuGan2008PhysA,XuGan2011CTP,LB2KHI2011,XuChen2009,
MRT2010EPL,XuChen2010,MRT2011CTP,MRT2011PLA,XuChen2011,MRT2011TAML,Review2012}.
For modeling and simulating high speed compressible flows, especially those with shocks, many attempts and considerable progress have been achieved
\cite{XuPan2007,XuGan2008CTP,XuGan2008PhysA,XuGan2011CTP,LB2KHI2011,XuChen2009,
MRT2010EPL,XuChen2010,MRT2011CTP,MRT2011PLA,XuChen2011,MRT2011TAML,Review2012}.

It should be pointed out that, up to now, most of LB models for compressible fluids are based on the Cartesian coordinate system. In many cases the flows show divergent, convergent, and/or rotational behaviors, for example, in
cylindrical or spherical devices. For such flow systems, LB models based on polar coordinates, cylindrical coordinates or spherical coordinates are more convenient and are less exposed to numerical errors. There have been a number of LB methods based on curvilinear coordinates or for axisymmetric cylindrical coordinate system. Early in 1992, Nannelli and Succi \cite{Nannelli1992} presented a general framework to extend the lattice Boltzmann equation to arbitrary lattice geometries. In this work, a finite-volume formulation of LB equation was given. Then some other finite-volume versions of the LB method were proposed for irregular meshes \cite{Succi1995,Amati1997,Peng1998,Peng1999,Ubertini2003}. In 1997 He and Doolen \cite{He1997} extended the LB method to apply to general curvilinear coordinate systems via using an interpolation-based strategy. In the following year Mei and Shyy \cite{Mei1998} developed a FDLB method in body-fitted curvilinear coordinates with non-uniform grids.
Later, Halliday et al \cite{Halliday2001} proposed a Polar Coordinate Lattice Boltzmann(PCLB) method for hydrodynamics. In 2005 Premnath and Abraham \cite{Premnath2005} presented a LB model for axisymmetric multiphase flows. In this work source terms were added to a two-dimensional standard LB equation for multiphase flows such that the emergent dynamics can be transformed into the axisymmetric cylindrical coordinate system. But all those LB methods work only for isothermal and nearly incompressible flows. In 2010 Asinari et al \cite{Asinari2010} formulated a LB scheme to analyze the radiative heat transfer problems in a participation medium, but did not take into account the effects of fluid flow. In 2011 Watari \cite{Watari2011} formulated a polar coordinate FDLB scheme to investigate the rotational flow problems in coaxial cylinders. This work presents valuable information on the LB application to the cylindrical system. However, this model works also only for subsonic flow systems. In the present work we extend the FDLB model based on polar coordinates to compressible flow systems with high Mach number so that it can be used to simulate flows with shock waves.

The rest of the paper is structured as follows. In section II we first briefly review the polar FDLB model by Watari, then present our contributions to the polar FDLB model. A hybrid scheme being similar to, but different from, the operator-splitting scheme is presented. In the combined scheme, the analytical solution for the temporal evolution and the Modified Warming-Beam (MWB) scheme for the convection behavior are used. Section III is for the validation and verification of the new LB model. In section IV we study the non-equilibrium characteristics of the system in two special cases related to shock wave passing material interfaces. The method to qualitatively recover the actual distribution function is illustrated. Section V concludes the present paper.

\section{Polar FDLB model}

\subsection{Brief review of Watari model}

\begin{figure}[tbp]
\center\includegraphics*[bbllx=0pt,bblly=0pt,bburx=595pt,bbury=204pt,angle=0,width=0.9\textwidth]
{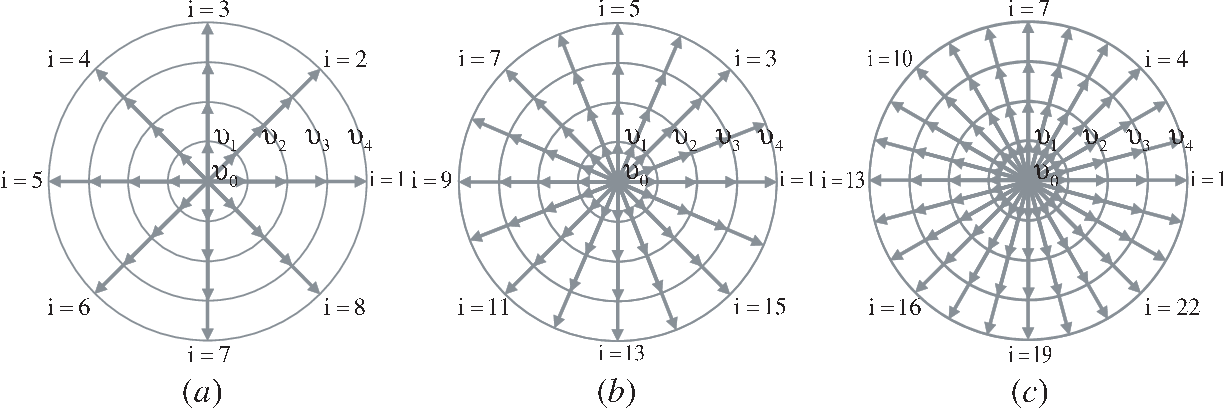}
\caption{Sketches of DVM with fixed $N_k=4$ and various values of $N_i$. (a)$N_i=8$. (b)$N_{i}$=16. (c)$N_i=24$.}
\label{Fig01}
\end{figure}

Below is a general description of the two-dimensional FDLB thermal model \cite{Watari2011}, which is applicable to both rectangular cartesian coordinate system and polar coordinate system. The evolution of the distribution
function $f_{ki}$ with the Bhatanger-Gross-Krook approximation \cite{bgk_1954} reads,
\begin{equation}
\frac{\partial f_{ki}}{\partial t}+\mathbf{v}_{ki}\cdot \nabla f_{ki}=-\frac{1}{\tau }(f_{ki}-f_{ki}^{eq})
\label{FDLBM}
\end{equation}
where $f_{ki}$ ($f_{ki}^{eq}$) is the discrete (equilibrium) distribution function; $\tau $ is the relaxation time determining the speed of approaching equilibrium; $\mathbf{v}_{ki}$ is the discrete velocity which will be defined below. The original Discrete-Velocity-Model (DVM) by Watari and Tsutahara is composed of ($N_{k}+1$) groups of discrete velocities. The $k$-th group has the size $v_{k}$. The first group has one component and each of the other group has $N_{i}$ components distributed in $N_{i}$ directions. Mathematically, the DVM can be written as:
\begin{equation}
\mathbf{v}_{ki}=\sum_{\alpha }v_{ki\alpha }\mathbf{e}_{\alpha }=v_{kix}
\mathbf{e}_{x}+v_{kiy}\mathbf{e}_{y}
\text{,}
\label{DVM1}
\end{equation}
where $\mathbf{e}_{x}$ and $\mathbf{e}_{y}$ are unit vectors in two-dimensional rectangular cartesian coordinate system, $v_{kix}=v_{k}\cos[2\pi (i-1)/N_{i}]$, $v_{kiy}=v_{k}\sin[2\pi (i-1)/N_{i}]$, $k=0$,$1$,$2$,$\cdots $,$N_{k}$, and $i=1$,$2$,$\cdots $,$N_{i}$. In this work we discuss the polar coordinate FDLB model for fixed $N_{k}=4$ and flexible $N_{i}$. The sizes of discrete velocities are chosen as $v_{0}=0$, $v_{1}=1$, $v_{2}=2.92$, $v_{3}=2.99$, $v_{4}=4.49$. The sketches of the DVM for cases of $N_{i}=8$, $N_{i}=16$, $N_{i}=24$ are shown in Fig.\ref{Fig01}.
%%%%%%%%%%%%%%%%%%%%%%%%%%%%%%%%%%%%%%%%
%%%%%%%%%%%%%%%%%%%%%%%%%%%%%%%%%%%%%%%%%

It's easy to prove that this DVM with $N_{i}=8$ has at least up to seventh rank isotropy\cite{XuGan2008PhysA}. The macroscopic quantities are defined as
\begin{equation}
\rho =\sum_{ki}f_{ki}^{eq}=\sum_{ki}f_{ki}\text{,}
\label{moment1}
\end{equation}
\begin{equation}
\rho \mathbf{u}=\sum_{ki}f_{ki}^{eq}\mathbf{v}_{ki}=\sum_{ki}f_{ki}\mathbf{v}_{ki}\text{,}
\label{moment2}
\end{equation}
\begin{equation}
\rho E=\sum_{ki}\frac{1}{2}f_{ki}^{eq}(\mathbf{v}_{ki}-\mathbf{u})\cdot (%
\mathbf{v}_{ki}-\mathbf{u})=\sum_{ki}\frac{1}{2}f_{ki}(\mathbf{v}_{ki}-%
\mathbf{u})\cdot (\mathbf{v}_{ki}-\mathbf{u})\text{.}
\label{moment3}
\end{equation}
Here $\rho $, $\mathbf{u}$ ($=u_{r}\mathbf{e}_{r}+u_{\theta }\mathbf{e}_{\theta }=
u_{x}\mathbf{e}_{x}+u_{y}\mathbf{e}_{y}$), $P$ ($=\rho T$), $E$($=T/(\gamma -1)$)
are the hydrodynamic density, flow velocity, pressure and internal kinetic energy per unit mass, respectively; $T$ is the temperature and $\gamma (=2)$ is the specific-heat ratio. Other velocity moments that the local equilibrium distribution function has to satisfy are:
\begin{equation}
\sum_{ki}f_{ki}^{eq}\mathbf{v}_{ki}\mathbf{v}_{ki}=\rho (E\mathbf{I}+\mathbf{uu})\text{,}
\label{moment4}
\end{equation}
\begin{equation}
\sum_{ki}f_{ki}^{eq}\mathbf{v}_{ki}\mathbf{v}_{ki}\mathbf{v}_{ki}=\rho
\lbrack E(\mathbf{u}_{\alpha}\mathbf{e}_{\beta}\mathbf{e}_{\gamma}\delta_{\beta \gamma}+\mathbf{e}_{\alpha}\mathbf{u}_{\beta}\mathbf{e}_{\gamma}\delta _{\gamma \alpha}
+\mathbf{e}_{\alpha}\mathbf{e}_{\beta}\mathbf{u}_{\gamma}\delta _{\alpha \beta})+\mathbf{uuu}]\text{,}
\label{moment5}
\end{equation}
\begin{equation}
\sum_{ki}\frac{1}{2}f_{ki}^{eq}\mathbf{v}_{ki}\cdot \mathbf{v}_{ki}\mathbf{v}_{ki}
=\rho \mathbf{u} (2E+\frac{1}{2} \mathbf{u} \cdot \mathbf{u})\text{,}
\label{moment6}
\end{equation}
\begin{equation}
\sum_{ki}\frac{1}{2}f_{ki}^{eq}\mathbf{v}_{ki}\cdot \mathbf{v}_{ki}\mathbf{v}_{ki}\mathbf{v}_{ki}
=\rho (2E+\frac{1}{2}\mathbf{u\cdot u})(E\mathbf{I}+\mathbf{uu})
\text{,}
\label{moment7}
\end{equation}
where $\mathbf{I}$ is the unit tensor, $\mathbf{v}_{ki}\mathbf{v}_{ki}$ and $\mathbf{uu}$ are double dyadics, $\mathbf{v}_{ki}\mathbf{v}_{ki}\mathbf{v}_{ki}$ and $\mathbf{uuu}$ are triple dyadics.

The equilibrium distribution function $f_{ki}^{eq}$ is computed by,
\begin{eqnarray}
f_{ki}^{eq} &=&\rho F_{k}[(1-\frac{u^{2}}{2E}+\frac{u^{4}}{8E^{2}})+
\frac{v_{ki\varepsilon }u_{\varepsilon }}{E}(1-\frac{u^{2}}{2E})+
\frac{v_{ki\varepsilon }v_{ki\pi }u_{\varepsilon }u_{\pi }}{2E^{2}}(1-\frac{u^{2}}{2E})
\notag \\
&&+\frac{v_{ki\varepsilon }v_{ki\pi }v_{ki\vartheta }u_{\varepsilon }u_{\pi
}u_{\vartheta }}{6E^{3}}+\frac{v_{ki\varepsilon }v_{ki\pi }v_{ki\vartheta
}v_{ki\xi }u_{\varepsilon }u_{\pi }u_{\vartheta }u_{\xi }}{24E^{4}}]
\label{feq}
\end{eqnarray}
with the weighting coefficients calculated in the following way,
\begin{subequations}
\begin{eqnarray}
F_{k} &=&\frac{1}{v_{k}^{2}(v_{k}^{2}-v_{k+1}^{2})(v_{k}^{2}-v_{k+2}^{2})(v_{k}^{2}-v_{k+3}^{2})}
[B_{4}E^{4}+B_{3}(v_{k+1}^{2}+v_{k+2}^{2}+v_{k+3}^{2})E^{3}
\notag \\
&&+B_{2}(v_{k+1}^{2}v_{k+2}^{2}+v_{k+2}^{2}v_{k+3}^{2}+v_{k+3}^{2}v_{k+1}^{2})E^{2}+B_{1}v_{k+1}^{2}v_{k+2}^{2}v_{k+3}^{2}E]
\text{,} \\
F_{0} &=&1-B_{0}(F_{1}+F_{2}+F_{3}+F_{4})
\text{,}
\end{eqnarray}
\end{subequations}
where the suffixes $\{k+l\}=\mod\{k+l, 4\}$, $l=0,1,2,3$, and the function $\mod\{a,b\}$ is defined as
\begin{equation}
\mod\{a,b\}=\left\{
\begin{array}{ccc}
a & \text{\ if } & a\leq b \\
a-b & \text{\ if } & a>b
\end{array}
\right. \text{.}
\end{equation}
%%%%%%%%%%%%%%%%%%%%%%%%%%%%%%%%%%%%%%%%%
\begin{center}
\begin{table}[tbp]
\caption{Coefficients $B_{0}$, $B_{4}$, $B_{3}$, $B_{2}$, $B_{1}$ for each model}
\begin{tabular}{ccccccc}
\hline\hline
model & ~~$N_i$~ & ~$B_{0}$~ & ~$B_{4}$~ & ~$B_{3}$~ & ~$B_{2}$~ & ~$B_{1}$~\\
\hline
Octagon & $8$ & $8$ & $48$ & $-6$ & $1$ & $-\frac{1}{4}$ \\
Double Octagon & $16$ & $16$ & $24$ & $-3$ & $\frac{1}{2}$ & $-\frac{1}{8}$ \\
Triple Octagon & $24$ & $24$ & $16$ & $-2$ & $\frac{1}{3}$ & $-\frac{1}{12}$\\
\hline\hline
\end{tabular}
\end{table}
\end{center}
%%%%%%%%%%%%%%%%%%%%%%%%%%%%%%%%%%%%%%%%%
The coefficients $B_{0}$, $B_{4}$, $B_{3}$, $B_{2}$, $B_{1}$ for each model are summarized in Table I.

Via the Chapman-Enskog expansion it is easy to find that this model presents the same results as the following Navier-Stokes equations
\begin{equation}
\frac{\partial \rho }{\partial t}+\nabla \cdot (\rho \mathbf{u})=0\text{,}
\label{NS_1}
\end{equation}
\begin{equation}
\frac{\partial (\rho \mathbf{u})}{\partial t}+\nabla \cdot (P\mathbf{I}
+\rho \mathbf{uu})+\nabla \cdot \lbrack \mu (\nabla \cdot \mathbf{u})\mathbf{I}
-\mu (\nabla \mathbf{u})^{T}-\mu \nabla \mathbf{u}]=0 \text{,}
\label{NS_2}
\end{equation}
\begin{gather}
\frac{\partial }{\partial t}(\rho E+\frac{1}{2}\rho u^{2})+\nabla \cdot
\lbrack \rho \mathbf{u}(E+\frac{1}{2}u^{2}+\frac{P}{\rho })]  \notag \\
-\nabla \cdot \lbrack \kappa ^{^{\prime }}\nabla E+\mu \mathbf{u}\cdot
(\nabla \mathbf{u})-\mu \mathbf{u}(\nabla \cdot \mathbf{u})+\frac{1}{2}\mu
\nabla u^{2}]=0 \text{,}
\label{NS_3}
\end{gather}
in the hydrodynamic limit, where $\mu$(=$P\tau$) and $\kappa ^{^{\prime }}$($=2P\tau$) are viscosity and heat conductivity, respectively.

\subsection{Our contribution}

In the system under consideration, if the collision term is directly treated with FD scheme, a stiff problem may occur; if the convection term is simply treated with FD scheme, the unphysical oscillations will be caused around strong discontinuity. In order to avoid or mitigate the two problems, we propose a new FD scheme based on a similar idea as the operator splitting scheme.

\subsubsection{Hybrid scheme}

Equation \eqref{FDLBM} could be written in the following scalar form in polar coordinates
\begin{equation}
\frac{\partial f_{ki}}{\partial t}+v_{kir}\frac{\partial f_{ki}}{\partial r}+\frac{1}{r}v_{ki\theta }\frac{\partial f_{ki}}{\partial \theta }=-\frac{1}{\tau }(f_{ki}-f_{ki}^{eq}) \text{.}
\label{PolarLBE1}
\end{equation}
The two-dimensional FDLB Eq.\eqref{PolarLBE1} can be decomposed into the following one-dimensional form
\begin{equation}
\left\{
\begin{array}{c}
\frac{\partial f_{ki}}{\partial t}=-\frac{1}{\tau }(f_{ki}-f_{ki}^{eq}) \\
\frac{\partial f_{ki}}{\partial t}+v_{kir}\frac{\partial f_{ki}}{\partial r}=0 \\
\frac{\partial f_{ki}}{\partial t}+\frac{1}{r}v_{ki\theta }\frac{\partial f_{ki}}{\partial \theta }=0
\text{.}
\end{array}
\right.
\label{operator_splitting}
\end{equation}

\subsubsection{Analytic solution for temporal evolution}

The first subequation in Eq.\eqref{operator_splitting} has a traditional discrete solution in the following form
\begin{equation}
f_{ki}^{t+\Delta t}=f_{ki}^{t}-\frac{\Delta t}{\tau }(f_{ki}^{t}-f_{ki}^{eq}) \text{.}
\label{operator1_old}
\end{equation}
In fact the subequation can be given an analytical solution dynamically as below
\begin{equation}
f_{ki}^{t+dt}=f_{ki}^{eq}+(f_{ki}^{t}-f_{ki}^{eq})\exp (-\frac{dt}{\tau }) \text{.} \label{operator1_new}
\end{equation}

\subsubsection{MWB scheme for spatial evolution}

The last two subequations in Eq.\eqref{operator_splitting} can be written uniformly as
\begin{equation}
\begin{array}{c}
\frac{\partial \psi }{\partial t}+a\frac{\partial \psi }{\partial \xi }=0 \\
\psi \rightarrow f_{ki} \\
\xi \rightarrow r\ or\ \theta \\
a\rightarrow v_{kir}\ or\ \frac{1}{r}v_{ki\theta } \text{.}
\end{array}
\label{Euler_1order}
\end{equation}
Since $v_{kir}$ is a constant and $r$ can also be regarded as a constant when consider the last subequation of Eq.\eqref{operator_splitting}, we can further obtain
\begin{equation}
\frac{\partial ^{2}\psi }{\partial t^{2}}-a^{2}\frac{\partial ^{2}\psi }{\partial \xi ^{2}}=0  \text{.}  \label{Euler_2order}
\end{equation}
By introducing the symbol, $\psi (\xi _{j},t_{n})=\psi _{j}^{n}$, and performing the Taylor expansion, we get
\begin{equation}
\psi _{j}^{n+1}=\psi _{j}^{n}-a\Delta t(\frac{\partial \psi }{\partial \xi })_{j}^{n}+\frac{1}{2}a^{2}\Delta t^{2}(\frac{\partial ^{2}\psi }{\partial\xi ^{2}})_{j}^{n}+O(\Delta t^{3})\text{.}  \label{Taylor expansion}
\end{equation}
The derivatives about $\xi _{j}$ in Eq.\eqref{Taylor expansion} are all calculated with the second order upwind scheme,
\begin{eqnarray}
(\frac{\partial \psi }{\partial \xi })_{j}^{n} &=&\left\{
\begin{array}{ccc}
\frac{3\psi _{j}^{n}-4\psi _{j-1}^{n}+\psi _{j-2}^{n}}{2\Delta \xi }+O(\Delta \xi ^{2}) & \text{\ if } & a\geq 0 \\
-\frac{3\psi _{j}^{n}-4\psi _{j+1}^{n}+\psi _{j+2}^{n}}{2\Delta \xi }+O(\Delta \xi ^{2}) & \text{\ if } & a<0
\end{array}
\right.  \label{upwind scheme_1} \\
(\frac{\partial ^{2}\psi }{\partial \xi ^{2}})_{j}^{n} &=&\left\{\begin{array}{ccc}
\frac{\psi _{j}^{n}-2\psi _{j-1}^{n}+\psi _{j-2}^{n}}{\Delta \xi ^{2}}%
+O(\Delta \xi ) & \text{\ \ \ \ \ \ if } & a\geq 0 \\
\frac{\psi _{j}^{n}-2\psi _{j+1}^{n}+\psi _{j+2}^{n}}{\Delta \xi ^{2}}%
+O(\Delta \xi ) & \text{\ \ \ \ \ \ if } & a<0
\end{array}
\right.
\label{upwind scheme_2}
\end{eqnarray}
Thus, from Eq.\eqref{Taylor expansion} we get the well-known Warming-Beam Scheme,
\begin{equation}
\psi _{j}^{n+1}=\left\{
\begin{array}{ccc}
\psi _{j}^{n}-C(\psi _{j}^{n}-\psi _{j-1}^{n})-\frac{1}{2}C(1-C)(\psi
_{j}^{n}-2\psi _{j-1}^{n}+\psi _{j-2}^{n}) & \text{if} & C\geq 0 \\
\psi _{j}^{n}-C(\psi _{j+1}^{n}-\psi _{j}^{n})+\frac{1}{2}C(1+C)(\psi
_{j}^{n}-2\psi _{j+1}^{n}+\psi _{j+2}^{n}) & \text{if} & C<0%
\end{array}
\right.
\label{WB_Scheme_1}
\end{equation}
where the higher order tiny quantities have been omitted and $C$($=a\Delta t/\Delta \xi$) is the Courant-number. The stability condition requires $\left\vert C\right\vert \leq 2$.

In this work we modify the Warming-Beam scheme. Firstly, Eq.\eqref{WB_Scheme_1} is changed into the following form
\begin{equation}
\psi _{j}^{n+1}=\psi _{j}^{n}-[C+\frac{1}{2}C(1-\left\vert C\right\vert)(1-\eta )]\delta
\label{WB_Scheme_2}
\end{equation}
where
\begin{equation}
\delta =\left\{
\begin{array}{ll}
\psi _{j}^{n}-\psi _{j-1}^{n} & \text{\ if }C\geq 0 \text{,} \\
\psi _{j+1}^{n}-\psi _{j}^{n} & \text{\ if }C<0 \text{,}
\end{array}
\right.
\eta =\left\{
\begin{array}{cc}
\frac{\psi _{j-1}^{n}-\psi _{j-2}^{n}}{\psi _{j}^{n}-\psi _{j-1}^{n}} & \text{\ if }C\geq 0 \text{,} \\
\frac{\psi _{j+2}^{n}-\psi _{j+1}^{n}}{\psi _{j+1}^{n}-\psi _{j}^{n}} & \text{\ if }C<0 \text{.}
\end{array}
\right.
\end{equation}
In principle, any linear difference scheme causes dispersion and dissipation problems. Hence, no linear difference scheme is suitable for solving strong discontinuity problems. Using non-linear difference scheme is necessary. The simplest solution is to use piecewise linear difference scheme. This is the reason why we introduce a switch function $S(\eta )$ into Eq.\eqref{WB_Scheme_2}, i.e.,
\begin{equation}
\psi _{j}^{n+1}=\psi _{j}^{n}-[C+\frac{1}{2}C(1-\left\vert C\right\vert)(1-S(\eta ))(1-\eta )]\delta \text{.}
\label{MWB Scheme}
\end{equation}
To make the scheme monotonous in space, we require
\begin{equation}
0\leq \varphi (C)\leq 1,
\label{varphi_0}
\end{equation}
where $\varphi (C)$\ is a quadratic polynomial function
\begin{equation}
\varphi (C)=\left\vert C\right\vert +\frac{1}{2}\left\vert C\right\vert
(1-\left\vert C\right\vert )(1-S(\eta ))(1-\eta ) \text{.}
\label{varphi_1}
\end{equation}
From Eqs.\eqref{varphi_0}-\eqref{varphi_1} we get
\begin{equation}
\left\{
\begin{array}{c}
\varphi (0)=0 \\
0\leq \varphi (C)\leq 1 \\
\varphi (1)=\varphi (-1)=1
\text{.}
\end{array}
\right.
\label{varphi_2}
\end{equation}
Equation \eqref{varphi_1} can be written as
\begin{equation}
g(x)=x+\frac{1}{2}x(1-x)\alpha \text{,}
\label{varphi2b}
\end{equation}
where $x=\left\vert C\right\vert $, $g(x)=\varphi (C)$, $\alpha=(1-S(\eta ))(1-\eta )$. Thus, Eq.\eqref{varphi_2} becomes
\begin{equation}
\left\{
\begin{array}{c}
g(0)=0 \text{,} \\
0\leq g(x)\leq 1 \text{,}\\
g(1)=1 \text{.}
\end{array}
\right.
\label{varphi2c}
\end{equation}
Equation \eqref{varphi2b} describes a parabola which has an extremum value at
\begin{equation}
x_{e}=\frac{1}{2}+\frac{1}{\alpha }\text{.}  \label{varphi2d}
\end{equation}
To satisfy all the three conditions in Eq.\eqref{varphi2c}, we require
\begin{equation}
x_{e}\geq 1\text{ or }x_{e}\leq 0  \label{varphi2e}
\end{equation}
From the conditions in Eq.\eqref{varphi2e} we have $|\alpha |\leq 2$, i.e.,
\begin{equation}
\left\vert (1-S(\eta ))(1-\eta )\right\vert \leq 2\text{.}
\end{equation}
So we choose
\begin{equation}
S(\eta )=\frac{\left\vert \eta \right\vert -1}{\left\vert \eta \right\vert +1} \text{.}
\end{equation}
To this step, we have got a new conservative monotonous scheme with second-order accuracy. It should be pointed out that, besides the lattice Boltzmann equation, the MWB scheme also works for simulating hydrodynamic equations.

\subsubsection{Combined scheme for the LB evolution}

By composing the solutions of the three subequations in Eq.\eqref{operator_splitting}, we get the combined scheme for the LB evolution,
\begin{equation}
\begin{array}{c}
f_{ki}^{t+\Delta t}
=f_{ki}^{eq}+(f_{ki}^{t}-f_{ki}^{eq})\exp (-\frac{\Delta t}{\tau}) \\
-[C_{r}+\frac{1}{2}C_{r}(1-\left\vert C_{r}\right\vert )(1-S(\eta_{r}))(1-\eta _{r})]\delta _{r} \\
-[C_{\theta }+\frac{1}{2}C_{\theta }(1-\left\vert C_{\theta }\right\vert)(1-S(\eta _{\theta }))(1-\eta _{\theta })]\delta _{\theta }
\end{array}
\label{combined_scheme}
\end{equation}
with
\begin{equation}
\begin{array}{ccc}
C_{r}=v_{kir}\frac{\Delta t}{\Delta r} & \text{ , } &
C_{\theta }=\frac{1}{r}v_{ki\theta }\frac{\Delta t}{\Delta \theta }
\end{array}
\notag
\end{equation}
\begin{equation}
\begin{array}{c}
\delta _{r}=\left\{
\begin{array}{ccc}
f_{ki}(i_r,i_\theta )-f_{ki}(i_r-1,i_\theta ) & \text{ if } & v_{kir}\geq 0 \\
f_{ki}(i_r+1,i_\theta )-f_{ki}(i_r,i_\theta ) & \text{\ if } & v_{kir}<0
\end{array}
\right. \\
\delta _{\theta }=\left\{
\begin{array}{ccc}
f_{ki}(i_r,i_\theta )-f_{ki}(i_r,i_\theta -1) & \text{\ if } & v_{ki\theta }\geq 0 \\
f_{ki}(i_r,i_\theta +1)-f_{ki}(i_r,i_\theta ) & \text{\ if } & v_{ki\theta }<0
\end{array}
\right.
\end{array}
\notag
\end{equation}
\begin{equation}
\begin{array}{c}
\eta _{r}=\left\{
\begin{array}{ccc}
\frac{f_{ki}(i_r-1,i_\theta )-f_{ki}(i_r-2,i_\theta )}{f_{ki}(i_r,i_\theta)-f_{ki}(i_r-1,i_\theta )} & \text{ if } & v_{kir}\geq 0 \\
\frac{f_{ki}(i_r+2,i_\theta )-f_{ki}(i_r+1,i_\theta )}{f_{ki}(i_r+1,i_\theta)-f_{ki}(i_r,i_\theta )} & \text{ if } & v_{kir}<0
\end{array}
\right. \\
\eta _{\theta }=\left\{
\begin{array}{ccc}
\frac{f_{ki}(i_r,i_\theta -1)-f_{ki}(i_r,i_\theta -2)}{f_{ki}(i_r,i_\theta)-f_{ki}(i_r,i_\theta -1)} & \text{ if } & v_{ki\theta }\geq 0 \\
\frac{f_{ki}(i_r,i_\theta +2)-f_{ki}(i_r,i_\theta +1)}{f_{ki}(i_r,i_\theta+1)-f_{ki}(i_r,i_\theta )} & \text{ if } & v_{ki\theta }<0
\end{array}
\right.
\end{array}
\notag
\end{equation}
where, $i_r$ and $i_\theta $ are indexes of the coordinate. The combined scheme has first-order accuracy as a whole, see appendix. The combined scheme is different from the Strang splitting scheme used in \cite{Dellar2013}. Compared with the latter, the present scheme is simpler. Meanwhile, numerical tests show that the numerical stability of the present scheme is almost the same as the one of the latter.

\subsection{Boundary conditions}

\begin{figure}[tbp]
\center\includegraphics*
[bbllx=0pt,bblly=0pt,bburx=595pt,bbury=285pt,angle=0,width=0.75\textwidth]{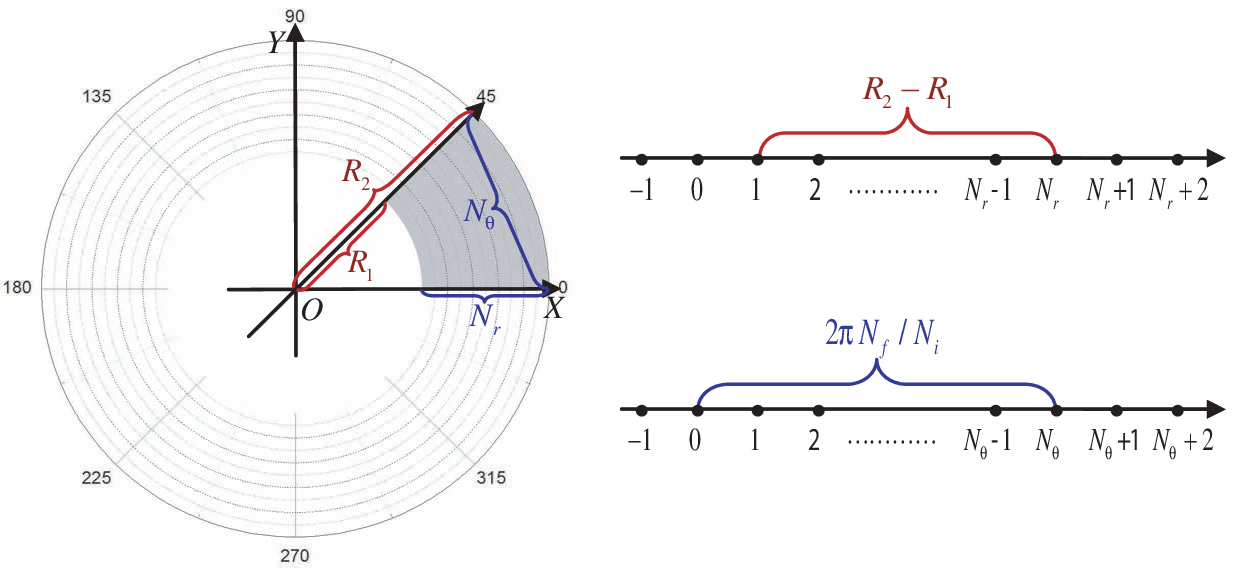}
\caption{Sketch for the whole system and the computational domain with lattice nodes.}
\label{Fig02}
\end{figure}

The physical domain under consideration is in an annular area with radii $R_{2}>R_{1}>0$. When the inner radius $R_{1}\rightarrow 0$, the annular area approximates to a circular one. If the annular physical domain is periodic and the period is $N_{i}$ along the circumferential direction, it can be sectioned into $N_{i}$ parts of sector, where $N_{i}$ is just the total number of the directions of discrete velocity in the DVM. In this case, we just pick out one part for calculations. If the period is $N_{i}/N_{f}$, we can pick out $N_{f}$ connected parts as computational domain, where $N_{f}$ is a positive integer. Thus, the computational domain is that with $R_{1}\leq r\leq R_{2}$
and $0<\theta \leq 2\pi N_{f}/N_{i}$. The computational domain has two boundaries in the radial direction and two in the circumferential direction. It is clear that periodic boundary conditions should be applied in the circumferential direction. On the other hand, the inner and outer boundaries in the radial direction should be treated specifically according to the specific situation under consideration. In this work we study the case with $N_{i}=8$ and $N_{f}=1$. Figure \ref{Fig02} shows a sketch for the computational domain.
%%%%%%%%%%%%%%%%%%%%%%%%%%%%%%%%%%%%%%%%
%%%%%%%%%%%%%%%%%%%%%%%%%%%%%%%%%%%%%%%%%

\subsubsection{Radial boundary condition}

Assume that the total number of radial nodes is $N_{r}$, radial increment is $\Delta r=(R_{2}-R_{1})/(N_{r}-1)$, and the radius is $r=R_{1}+(i_r-1)\times \Delta r$, $i_r=1,2,\cdots ,N_{r}$. In the case where the density can
be considered continuous around the boundaries, we can obtain the density values on the ghost nodes ($i_r=-1,0,N_{r}+1,N_{r}+2$) via linear interpolation scheme,
\begin{equation}
\rho (i_r,i_\theta )=\left\{
\begin{array}{ccc}
2\rho (i_r+1,i_\theta )-\rho (i_r+2,i_\theta ) & \text{ if } & i_r<1 \\
2\rho (i_r-1,i_\theta )-\rho (i_r-2,i_\theta ) & \text{ if } & i_r>N_{r}
\end{array}
\right.
\end{equation}
or an interpolation scheme with higher-order accuracy. The temperature can be calculated in a similar way. However, the determination of flow velocity depends on the specific situation under consideration. The simplest microscopic boundary condition is to assume that at each boundary node the system is in its thermodynamic equilibrium, i.e. $f_{ki}=f_{ki}^{eq}$. For the non-equilibrium microscopic boundary condition, the deviation from thermodynamic equilibrium, $f_{ki}-f_{ki}^{eq}$, can be calculated via the interpolation scheme\cite{Guo2002}.

\subsubsection{Azimuthal boundary condition}

\begin{figure}[tbp]
\center\includegraphics*
[bbllx=0pt,bblly=0pt,bburx=595pt,bbury=280pt,angle=0,width=0.75\textwidth]{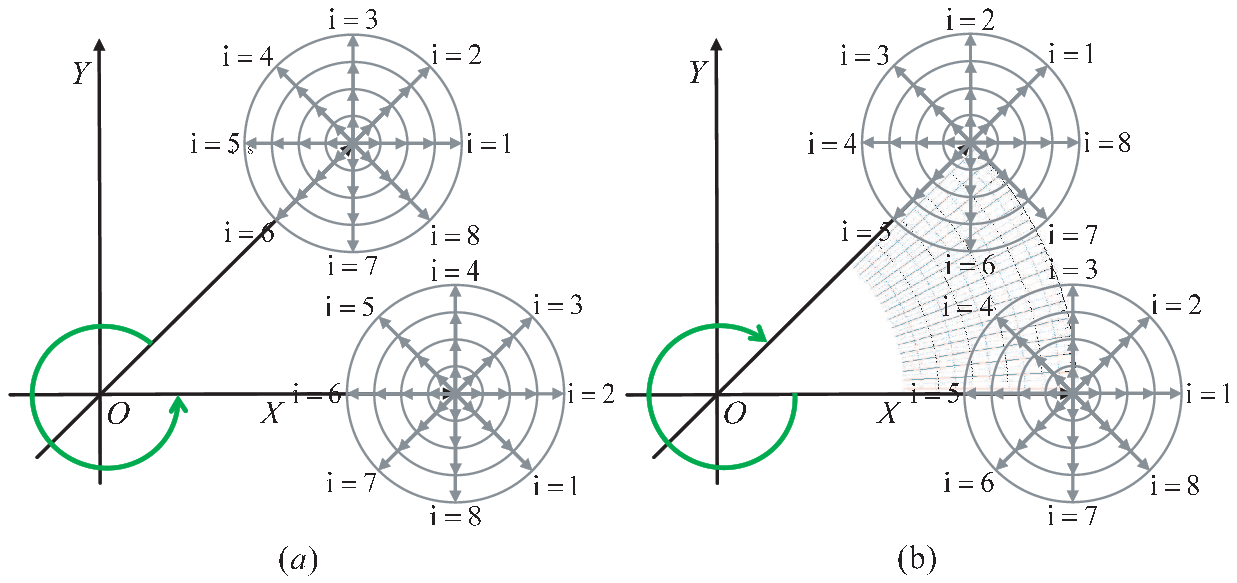}
\caption{(a) Rotation from the azimuthal boundary with $i_\theta=N_{\protect\theta }$ to the one with $i_\theta=0$.
(b) Rotation from the azimuthal boundary with $i_\theta=1$ to the one with $i_\theta=N_\theta+1$.}
\label{Fig03}
\end{figure}

Similarly, assume that the total number of azimuthal nodes is $N_{\theta }$, azimuthal increment is $\Delta \theta =2\pi N_{f}/(N_{i}N_{\theta })$, and the angle $\theta =i_\theta \times \Delta \theta $, $i_\theta=1,2,\cdots ,N_{\theta }$. The distribution functions on ghost nodes ($i_\theta=-1,0,N_{\theta }+1,N_{\theta }+2$) are computed in the following way
\begin{equation}
f(i_{r},i_{\theta },k,i)=\left\{
\begin{array}{ccc}
f(i_{r},N_{\theta }+i_{\theta },k,\mod\{i+N_{f},N_{i}\}) & \text{\ if } & i_{\theta }\leq 0 \\
f(i_{r},i_{\theta }-N_{\theta },k,\mod\{i-N_{f}+N_{i},N_{i}\}) & \text{\ if } & i_{\theta }>N_{\theta }
\end{array}
\right. \text{.}
\end{equation}
A schematic diagram for the case with $N_{i}=8$ and $N_{f}=1$ is referred to Fig.\ref{Fig03}. Figure (a) shows the way in which $f_{ki}(i_r,0)$ is given from $f_{ki}(i_r,N_{\theta })$ via rotation. Figure (b) shows the relation
between $f_{ki}(i_r,N_{\theta }+1)$\ and $f_{ki}(i_r,1)$ via rotation.
%%%%%%%%%%%%%%%%%%%%%%%%%%%%%%%%%%%%%%%%
%%%%%%%%%%%%%%%%%%%%%%%%%%%%%%%%%%%%%%%%%

\section{Validation and Verification}

\subsection{Performance on rotational flow}

We first consider the motion of a fluid between two coaxial cylinders, with radii $R_{1}$ and $R_{2}$, rotating about their axis with angular velocities $\omega _{1}$ and $\omega _{2}$. It should be pointed out that, the compressibility of the fluid is proportional to the Mach number squared. Our PCLB model is for compressible fluid and physically consistent with this behavior. In this test the Mach number is small, so we roughly consider the system as incompressible. Due to the rotational symmetry, we have $u_{r}=0$, $u_{\theta }=u_{\theta }(r)$, $P=P(r)$. For simplicity, we rewrite $u_{\theta }$ as $u$ in this test. The Navier-Stokes equation for incompressible flow in cylindrical polar coordinates gives the following two equations:
\begin{subequations}
\begin{eqnarray}
\frac{\partial P}{\partial r}-\rho u^{2}/r &=&0 \tt{,}\\
\mu (\frac{\partial ^{2}u}{\partial r^{2}}+\frac{1}{r}\frac{\partial u}{\partial r}-\frac{u}{r^{2}}) &=&0 \tt{.}
\end{eqnarray}
\label{RSF_NS}
\end{subequations}
The second has the following solution,
\begin{equation}
u=Ar+\frac{B}{r}
\label{RSF_u}
\end{equation}
where the constants $A$ and $B$ are found from the boundary conditions,
\begin{equation}
\left\{
\begin{array}{ccc}
u=R_{1}\omega _{1} & \text{ for } & r=R_{1} \\
u=R_{2}\omega _{2} & \text{ for } & r=R_{2}
\end{array}
\right. \text{.}
\label{RSF_boundary}
\end{equation}
As a result, we get the velocity distribution to be
\begin{equation}
u=\frac{\omega _{2}R_{2}^{2}-\omega _{1}R_{1}^{2}}{R_{2}^{2}-R_{1}^{2}}r+
\frac{R_{1}^{2}R_{2}^{2}(\omega _{1}-\omega _{2})}{R_{2}^{2}-R_{1}^{2}}\frac{1}{r}
\label{RSF}
\end{equation}
which is a non-slip Navier-Stokes solution and is independent of the
viscosity $\mu$.
%%%%%%%%%%%%%%%%%%%%%%%%%%%%%%%%%%%%%%%%
\begin{figure}[tbp]
\center\includegraphics*[bbllx=0pt,bblly=0pt,bburx=590pt,bbury=430pt,angle=0,width=0.45\textwidth]
{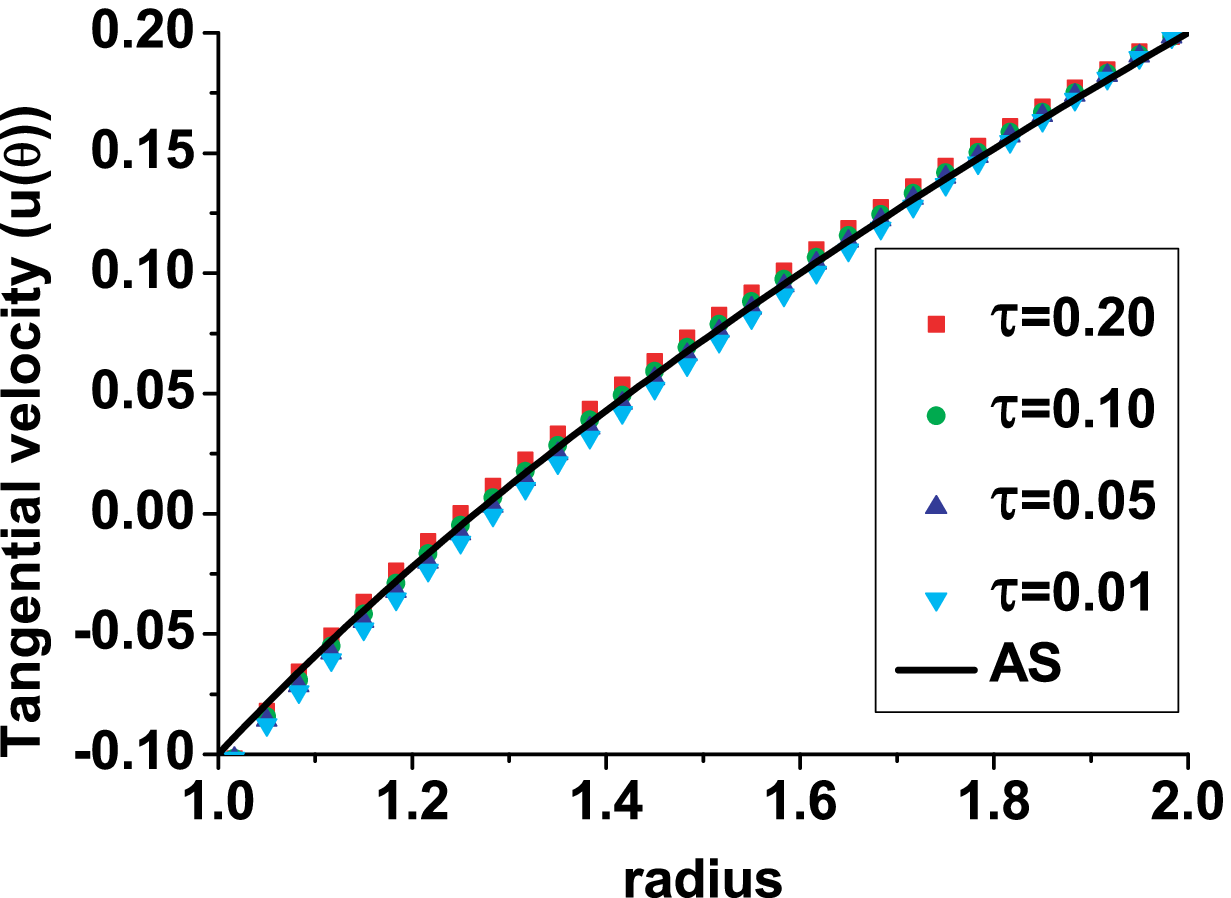}
\caption{Comparison of our PCLB results with analytical solution for the steady
rotational velocity $u_{\protect\theta }$ under various values of $\protect\tau $.}
\label{Fig04}
\end{figure}
%%%%%%%%%%%%%%%%%%%%%%%%%%%%%%%%%%%%%%%%%

Initially, the system is in its thermodynamic equilibrium with $\rho =1.0$, $T=1$ and $u_r =0$, $u_{\theta}=0$.
The other parameters are given as $R_{1}=1$, $R_{2}=2$, $\omega_{1}=-0.1$, $\omega _{1}=0.1$, $\Delta t=10^{-3}$,
$N_{r}\times N_{\theta }=100\times 20$. Figure \ref{Fig04} shows a comparison of the simulation results by
our PCLB model with the Analytic Solution (AS) for the final steady state, where the values of relaxation time
$\tau$ are $0.2$, $0.1$, $0.05$, and $0.01$, respectively. We can find that the simulation results have a satisfying agreement with AS. The slight mismatch is due to the weak compressibility of the fluid which is ignored in the analytical solution.
%%%%%%%%%%%%%%%%%%%%%%%%%%%%%%%%%%%%%%%%
\begin{figure}[tbp]
\center\includegraphics*[bbllx=0pt,bblly=0pt,bburx=595pt,bbury=296pt,angle=0,width=0.9\textwidth]{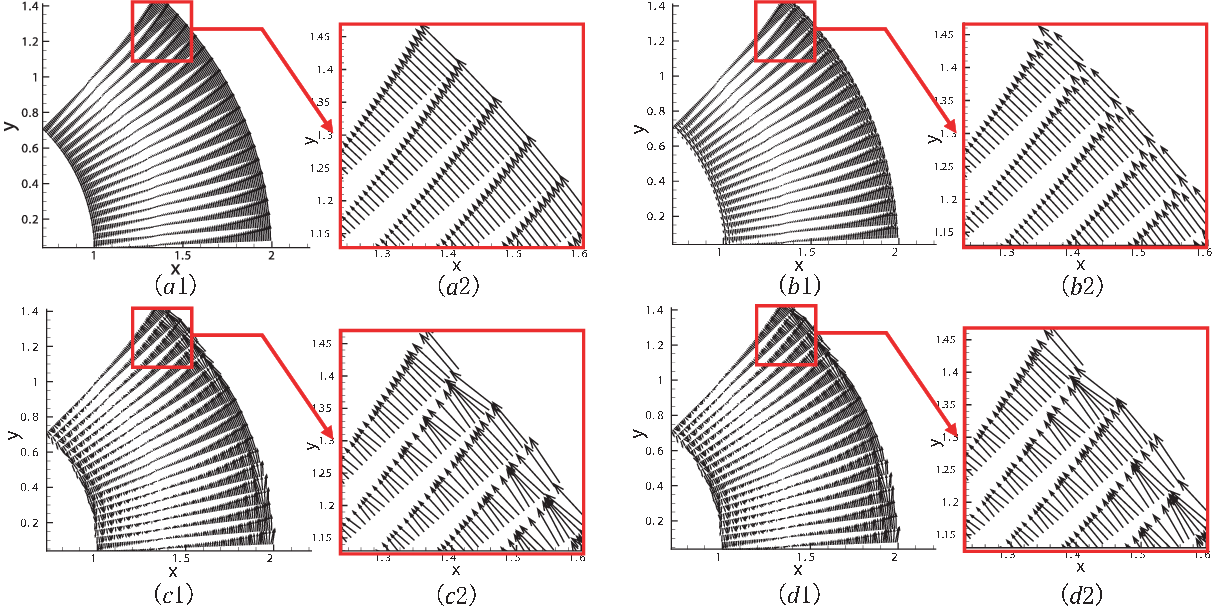}
\caption{The plane description of velocity field about the rotational flow at time $t=0.6$: (a) the MWB scheme; (b) Lax-Wendroff scheme; (c) the second order up wind scheme; (d) Warming-Beam scheme.}
\label{Fig05}
\end{figure}
%%%%%%%%%%%%%%%%%%%%%%%%%%%%%%%%%%%%%%%%%

To compare our MWB scheme with other FD schemes for the convection term, we show the simulation results of velocity field at time $t=0.6$ in Fig.\ref{Fig05}. The FD schemes used in Figs.(a1)-(d1) are our MWB scheme, the Lax-Wendroff scheme, the second order upwind scheme and the original Warming-Beam scheme, respectively. Figures (a2)-(d2) show the enlargements of the portions in the corresponding squares in Figs.(a1)-(d1). It is clear in Fig.(b2) that the Lax-Wendroff scheme brings artificial oscillations in the tangential component of flow velocities near the radial boundaries. From Figs.(c2) and (d2) we can find that both the second order upwind and the original Warming-Beam schemes bring artificial oscillations in the radial component of flow velocity in the whole computational domain. The simulation results from our new scheme have a satisfying agreement with theoretical analysis.

\subsection{Performance on discontinuity}

To check the performance of our new scheme on system with discontinuity, we consider a shock wave propagating outward in an annular system with radii $R_{1}$ and $R_{2}$. The physical quantities around shock front satisfy the following Hugoniot relations,
\begin{equation}
\left\{
\begin{array}{ccl}
\rho_{H}(u_{H}-D) & = & \rho _{0}(u_{0}-D) \\
P_{H}-P_{0} & = & \rho _{0}(D-u_{0})(u_{H}-u_{0}) \\
E_{H}-E_{0} & = & \frac{1}{2}(P_{H}+P_{0})(\frac{1}{\rho _{0}}-\frac{1}{\rho_{H}})
\end{array}
\right.
\label{Hugoniot}
\end{equation}
where $D$\ is the velocity of shock wave, the suffixes $H$ and $0$ indicate the shocked region and pre-shocked region,respectively.

The initial physical field is below
\begin{equation*}
\left\{
\begin{array}{lllll}
(\rho \text{,}u_{r}\text{,}u_{\theta }\text{,}P)_{inner} & = &
(1.58824\text{,}0.785674\text{,}0\text{,}2.66667) & \text{,} & R_{1}\leq r<R_{S} \\
(\rho \text{,}u_{r}\text{,}u_{\theta }\text{,}P)_{outer} & = &
(1\text{,}0\text{,}0\text{,}1) & \text{,} & R_{S}\leq r<R_{2}
\end{array}
\right.
\end{equation*}
where $R_{S}$ is the position of shock front. We choose $R_{1}=2000$, $R_{2}=2025$, $R_{S}=2001$, $D=2.12132$, $\tau =10^{-5}$, $\Delta t=10^{-5}$, $N_{r}\times N_{\theta }=250\times 3$.

%%%%%%%%%%%%%%%%%%%%%%%%%%%%%%%%%%%%%%%%
\begin{figure}[tbp]
\center\includegraphics*
[bbllx=0pt,bblly=0pt,bburx=585pt,bbury=240pt,angle=0,width=0.99\textwidth]{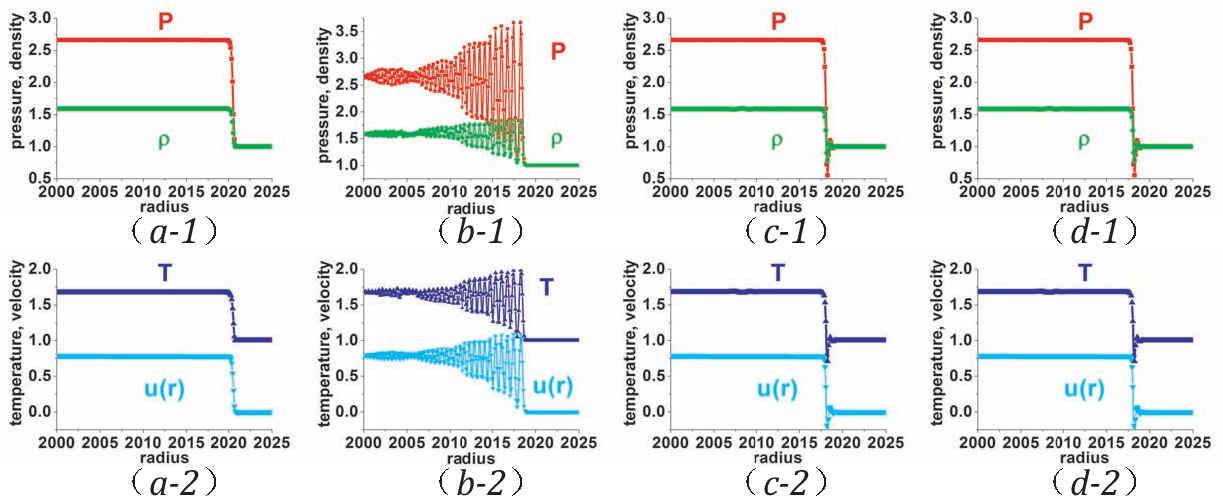}
\caption{Simulation results of physical quantities ($P$, $\rho $, $T$, $u_{r}$) along the radius with various schemes. From left to right, the four columns correspond to the MWB scheme, Lax-Wendroff scheme, the second order up wind scheme and Warming-Beam scheme.}
\label{Fig06}
\end{figure}
%%%%%%%%%%%%%%%%%%%%%%%%%%%%%%%%%%%%%%%%%
Figure \ref{Fig06} shows the simulation results of pressure $P$, density $\rho $, temperature $T$ and velocity $u_{r}$ along the radius at time $t=8$ using various schemes. From left to right, the four columns correspond to our MWB scheme, the Lax-Wendroff scheme, the second order upwind scheme and the original Warming-Beam scheme, respectively. The second column shows that the simulation results of physical quantities from the Lax-Wendroff scheme have strong unphysical oscillations in the shocked region. The third column shows that the second order upwind scheme brings unphysical \textquotedblleft overshoot\textquotedblright\ phenomena in physical quantities at the shock front. The fourth column shows that the original Warming-Beam has the same drawback as the second order upwind scheme. In contrast to the other three columns, the first column shows that the simulation results from our MWB scheme are much more accurate and physically reasonable.

\subsection{Simulation study on Richtmyer-Meshkov instability}

The Richtmyer-Meshkov (RM) instability takes place when a shock wave travels across an interface separating two kinds of fluids. For a two-dimensional rectangular system with a plane shock wave, several theoretical models have been proposed to describe the increase of the amplitude $A$. Roughly speaking, the increase rate of $A$ first shows a linear relationship with itself, i.e., $dA/dt=cA$, where $c$ is the increasing coefficient. In other words, the amplitude $A$ increases exponentially with time according to the relation, $A=A_{0}\exp (ct)$. When the time $t$ is very small, $\exp (ct)=1+ct$, $A=A_{0}+A_{0}ct$. It is clear that, at the very beginning, the amplitude $A$ linearly increases with time $t$. In the later time, the increasing coefficient $c$ itself is no longer a constant any more. Therefore, the later stage is generally referred to as the nonlinear increasing stage.

In 1960 Richtmyer \cite{Richtmyer1960} modified the linear theory of Taylor for Rayleigh-Taylor instability and proposed an impulsive model in the case of a reflected shock wave. The growth rate reads,
\begin{equation*}
\frac{dA}{dt}=k\Delta uA_{t} A_{1}\text{,}\qquad A_{1}=A_{0}(1-\frac{\Delta u}{D})
\end{equation*}
where $k$($=2\pi /\lambda$) is the wave number, $\Delta u$ is the velocity change of the material interface when shock passes, $A_{t}$ represents the post-shock Atwood number, $A_{1}$ is the post-shock amplitude, $A_{0}$ is the initial amplitude. $Cmpr$($=1-\Delta u/D$) is defined as compression ratio. In 1969 Meshkov \cite{Meshkov69} measured growth rate and found that it is only about one half of that predicted by the impulsive model. In 1992 Benjamin \cite{Benjamin92} got similar experiment results. In 1997 Zhang and Sohn \cite{Zhang1997} proposed a nonlinear model, using Pade approximation and asymptotic matching. The nonlinear model for two-dimensional system reads,
\begin{equation*}
\frac{dA}{dt}=\frac{v_{0}}{1+\zeta k^{2}v_{0}A_{1}t+\max[0,(kA_{1})^{2}-(A_{t})^{2}+0.5](kv_{0}t)^{2}}
\end{equation*}
where $v_{0}=k\Delta uA_{t} A_{1}$. This model is the growth rate of the perturbed material interface amplitude from early to late times in the cases of transition from light medium to heavy one ($\zeta =1$) and from heavy medium to light one ($\zeta =-1$)\cite{RMI_Review2002}.

In an annular system with radii $R_{1}=1.0$ and $R_{2}=2.0$, we study the RM instability in the following two cases: shocking from light to heavy media and shocking from heavy to light media. The initial sinusoidal perturbation, $r=R+A_{0}\times \cos (kR\theta )$, is applied to the density field, where $R$ is the mean position of the interface between the two media. Even though the system considered here shows two-dimensional geometrical effects, for the case where the perturbation wave length is small and the inner radius is large enough, the above theory for two-dimensional rectangular system with plane shock wave still works approximately.

\subsubsection{Shocking from light to heavy media}

We consider the case where a shock wave travels outward from a light medium to a heavy one with the velocity $D=2$. The initial physical field is given as below
\begin{equation*}
\left
\{
\begin{array}{lll}
(\rho ,u_{r},u_{\theta },P)_{inner} & = & (1.5,0.666667,0,2.33333) \\
(\rho ,u_{r},u_{\theta },P)_{middle} & = & (1,0,0,1) \\
(\rho ,u_{r},u_{\theta },P)_{outer} & = & (3,0,0,1)
\end{array}
\right.
\end{equation*}
where the subscripts $inner$ and $middle$ indicate shocked and pre-shocked regions of light medium, $outer$ represents outer region of heavy medium. The numerical values between the shocked and pre-shocked regions satisfy with the Hugoniot relations. We choose $A_0=0.02$, $R=1.2$, $k=20$, $\tau =\Delta t=10^{-5}$, $N_{r}\times N_{\theta }=1000\times 450$.

%%%%%%%%%%%%%%%%%%%%%%%%%%%%%%%%%%%%%%%%%%%%%%%%%%%%%%%%%%%%%%%%%%%%
\begin{figure}[tbp]
\center\includegraphics*[bbllx=36pt,bblly=256pt,bburx=557pt,bbury=451pt,angle=0,width=0.99\textwidth]{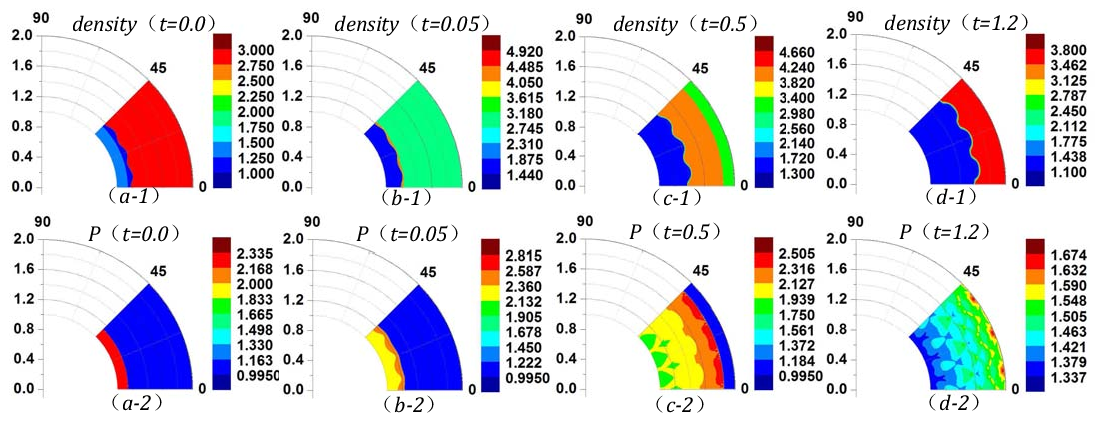}
\caption{Snapshots of RM instability for the case where the shock wave travels outwards from the light to heavy media. The density and pressure fields at the times, $t=0$, $t=0.05$, $t=0.5$, $t=1.2$, are shown from left to right, respectively.}
\label{Fig07}
\end{figure}
%%%%%%%%%%%%%%%%%%%%%%%%%%%%%%%%%%%%%%%%%%%%%%%%%%%%%%%%%%%%%%%%%%%%
Figure \ref{Fig07} shows the snapshots of the density and pressure fields. The first row is for the density fields. The second is for the pressure fields. From left to right, the four columns are for the times, $t=0$, $t=0.05$, $t=0.5$, $t=1.2$, respectively. When the shock wave passes the material interface, the perturbation amplitude $A$ in the density field first decreases significantly due to compression of the shock wave, see Figs.(a-1)-(b-1), then it begins to increase under the pressure gradient, see Figs.(b-1)-(d-1), where asymmetric structures at the two sides of the material interface eventually result in the occurrence of the bubbles in the light medium and spikes in the
heavy medium. The corresponding pressure fields shown in Figs. (a-2)-(d-2) present complementary information for understanding the evolution of the density field. It should be pointed out that the misalignment of pressure and density gradients promotes deformation of the material interface.

%%%%%%%%%%%%%%%%%%%%%%%%%%%%%%%%%%%%%%%%%%%%%%%%%%%%%%%%%%%%%%%%%%%%
\begin{figure}[tbp]
\center\includegraphics*[bbllx=0pt,bblly=0pt,bburx=595pt,bbury=128pt,angle=0,width=0.99\textwidth]{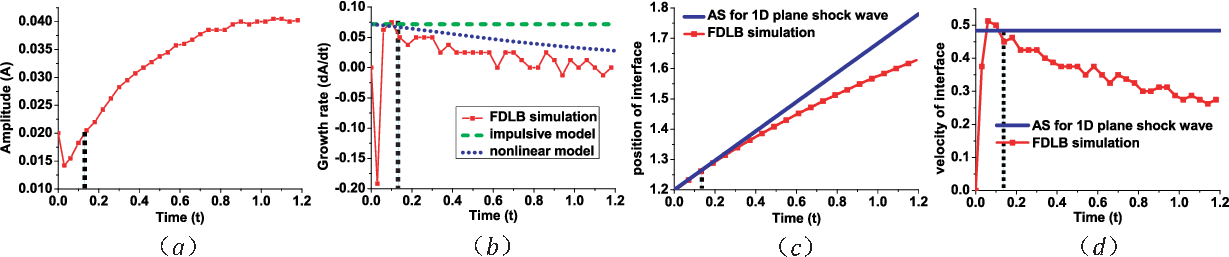}
\caption{Descriptions of the perturbed material interface in the evolution of RM instability for the case where a shock wave travels outwards from light to heavy media: (a) perturbation amplitude, (b) growth rate from various models, (c) radial position of material interface, (d) velocity of material interface. The vertical dashed line in each plot indicates the time when the perturbation amplitude recovers to its original value.}
\label{Fig08}
\end{figure}
%%%%%%%%%%%%%%%%%%%%%%%%%%%%%%%%%%%%%%%%%%%%%%%%%%%%%%%%%%%%%%%%%%%%
In order to draw some quantitative information to compare with the above theory, we show, from left to right in Fig.\ref{Fig08}, the amplitude, growth rate, mean position and velocity of material interface versus time, where $t=0$ is defined as the time when the shock wave meets with the perturbed material interface, and the amplitude is defined as one half of the maximum distance from the crest to trough of material interface. Figure (a) shows that
the evolution process of the perturbation amplitude can be roughly divided into three stages, the compression stage, the recovery stage and the further increasing stage. In the compression stage the amplitude drops rapidly to $A_{comp}=0.0143$ at the time $t_{comp}=0.03$. It recovers to its initial value at the time $t_{reco}=0.13$. A dashed vertical line is plotted in each of Figs.(a)-(d) to indicate the time, $t_{rec}$, when the perturbation amplitude recovers its original value. From the minimum value in Fig.(a), we get the compression ratio $Cmpr=0.0143/0.02=0.715$. However, the theoretical solution based on the initial conditions is $Cmpr=0.758$. The deviation of the simulation result from the theoretical one is about $6\%$. Figure (b) shows that the simulation result of growth rate of perturbation amplitude $A$ roughly agrees with those from theoretical models in the recovery stage. It should be pointed out that, after taking into account the two-dimensional effects existing in this polar coordinate test system but ignored by the theoretical models, the simulation result shows a satisfying agreement with the theories. In Figs.(c) and (d) we show the position and velocity of material interface by the red lines with scatters, respectively. To measure the divergent effects of the polar coordinate system, in Figs.(c) and (d) we present also the corresponding theoretical results for the simple one-dimensional problem where a plane shock wave passes the plane interface of two fluids. It is clear that the velocity of perturbed material interface is slower.
Physically, in the one-dimensional case, the shock wave does not result in transverse flow velocity, the material interface propagates in a constant velocity. While in the current case, two mechanisms are responsible for the decreasing of the propagation velocity of the material interface. Firstly, vortexes occur during the evolution of the RM instability. According to the energy conservation, the kinetic energy along the radial direction decreases. The second mechanism is related to the geometric effects of the polar coordinate system. With the propagation outwards, the area of the perturbed material interface becomes larger, the kinetic energy density decreases.

\subsubsection{Shocking from heavy to light media}

In the subsequent simulation, we choose $\rho _{inner}=1.5$, $\rho _{middle}=1$, $\rho_{outer}=0.5$, and other parameters are the same as those in the above simulation.

%%%%%%%%%%%%%%%%%%%%%%%%%%%%%%%%%%%%%%%%%%%%%%%%%%%%%%%%%%%%%%%%%%%%
\begin{figure}[tbp]
\center\includegraphics*[bbllx=0pt,bblly=0pt,bburx=595pt,bbury=189pt,angle=0,width=0.99\textwidth]{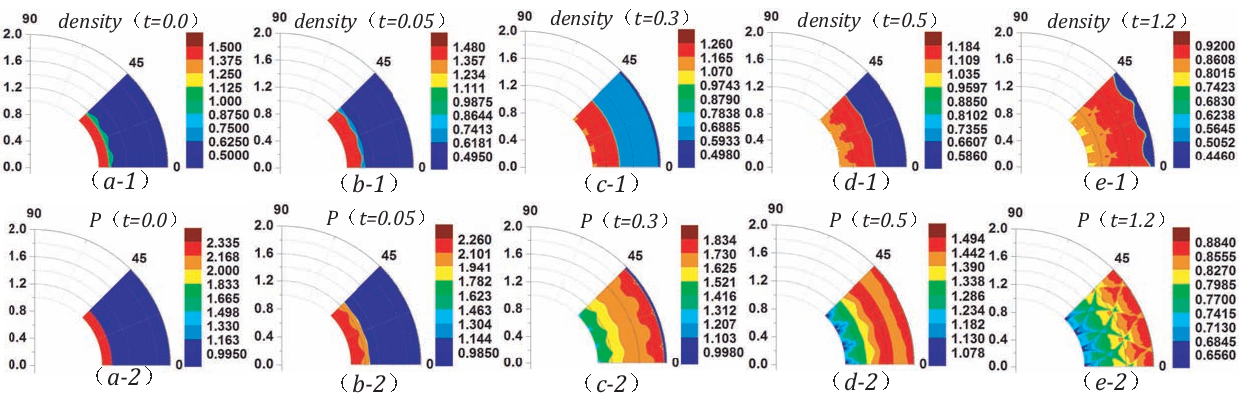}
\caption{Snapshots of RM instability for the case where the shock wave travels outwards from the heavy to light media. The density and pressure fields at the times, $t=0$, $0.05$, $0.3$, $0.5$ and $1.2$, are shown from left to right, respectively.}
\label{Fig09}
\end{figure}
%%%%%%%%%%%%%%%%%%%%%%%%%%%%%%%%%%%%%%%%%%%%%%%%%%%%%%%%%%%%%%%%%%%%
Figure \ref{Fig09} show the snapshots of density and pressure fields at times $t=0.0$, $0.05$, $0.3$, $0.5$ and $1.2$, respectively. The interface reversal phenomenon is obviously observed. When the shock wave passes the interface, a reflected rarefaction wave inward and a transmitted shock wave outward are generated. This stage
is known as the shock refraction stage. The pressure in heavy medium is smaller than that in the light medium near the crest of material interface. Driven by the pressure gradient, the perturbation amplitude decreases with the outward motion of the material interface. Then, the crest and trough of initial interface invert, the heavy and light fluids gradually penetrate into each other as time goes on, the light fluid \textquotedblleft falls" to form a bubble and the heavy fluid \textquotedblleft rises" to generate a spike.

%%%%%%%%%%%%%%%%%%%%%%%%%%%%%%%%%%%%%%%%%%%%%%%%%%%%%%%%%%%%%%%%%%%%
\begin{figure}[tbp]
\center\includegraphics*[bbllx=0pt,bblly=0pt,bburx=595pt,bbury=128pt,angle=0,width=0.99\textwidth]{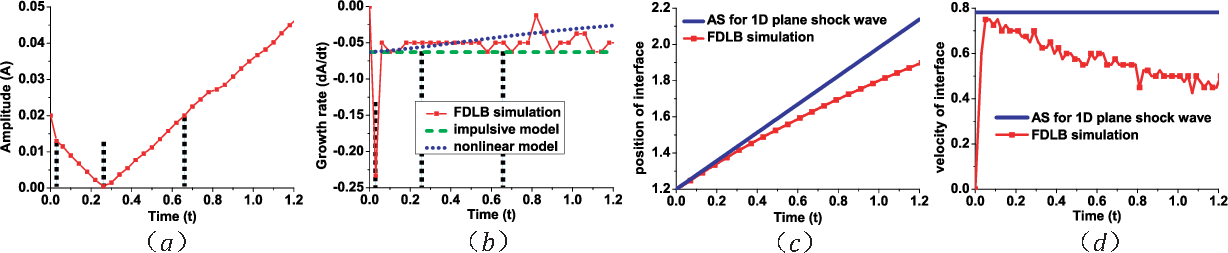}
\caption{Descriptions of the perturbed material interface in the evolution of RM instability for the case where shock wave travels outwards from heavy to light media: (a) perturbation amplitude, (b) growth rate from various models, (c) radial position of material interface, (d) velocity of material interface. Three vertical dashed lines are shown in each plot to guide the eyes for the compression, further compression, recovery and further increasing stages.}
\label{Fig10}
\end{figure}
%%%%%%%%%%%%%%%%%%%%%%%%%%%%%%%%%%%%%%%%%%%%%%%%%%%%%%%%%%%%%%%%%%%%
Figure \ref{Fig10} shows the descriptions of the perturbed material interface. Figure (a) shows the simulation results of perturbation amplitude. Figure (b) shows the growth rate, where the line with scatters is for the LB result, the dashed line is for numerical results from impulsive model and the dotted line is for the nonlinear model. Figures (c)-(d) show the mean interfacial position and the velocity of material interface along radius. Three vertical lines are shown in Figs.(a) and (b) to divide the evolution into four stages, i.e., the stages of initial compression, further compression, recovery and further increasing. The first guideline corresponds to the time, $t_{comp}=0.03$, when the amplitude is rapidly compressed to $A_{comp}=0.013$. The second one is for the time, $t_{zero}=0.27$, when the amplitude reaches zero. The last one indicates the time, $t_{reco}=0.66$, when the amplitude recovers to its initial value, $A_{init}=0.02$. From Fig.(a) we can get the compression ratio $Cmpr=0.013/0.02=0.65$. As a comparison, the theoretical solution is $Cmpr=0.61 $. It can be found in Fig.(b) that our simulation results are close to the results from the impulsive model and nonlinear model. Figures (c) and (d) show the same phenomena as those in Fig.\ref{Fig08}.

\subsection{Simulation study on Kelvin-Helmholtz instability}

To investigate the Kelvin-Helmholtz (KH) instability in an annular region with radii $R_{1}<R_{2}$, we set the initial physical field as below,
\begin{subequations}
\begin{eqnarray}
\rho (r) &=&\frac{{\rho _{inner}+\rho _{outer}}}{2}-\frac{{\rho_{inner}-\rho _{outer}}}{2}\tanh (\frac{r-R}{{D_{\rho }}})\text{,} \\
{\mathbf{u}}(r) &=&\frac{{\mathbf{u}_{inner}+\mathbf{u}_{outer}}}{2}-\frac{\mathbf{u}_{inner}{-\mathbf{u}_{outer}}}{2}\tanh (\frac{r-R}{{D_{u}}})\text{,}  \label{KH_u} \\
P(r) &=&P_{inner}=P_{outer}\text{,}
\end{eqnarray}
\label{KH_initial}
\end{subequations}
where $D_{\rho }$ and $D_{u}$ are the widths of density and velocity transition layers. $\rho _{inner}$, $\mathbf{u}_{inner}$ and $P_{inner}$ ($\rho _{outer}$, $\mathbf{u}_{outer}$ and $P_{outer}$) are the density, velocity and pressure of fluid near the inner (outer) cylinder, respectively. $R$ is the radial position of interface between two media. In order to trigger the KH rollup, the following perturbation of velocity in the $r$-direction,
\begin{equation}
u_{r}\mathbf{e}_{r}=u_{0}\mathbf{e}_{r}\sin (kR\theta )\exp (-\left\vert r-R\right\vert )\text{,}
\label{KH_perturbation}
\end{equation}
is added to the initial velocity field described by Eq.\eqref{KH_u}, where $u_{0}$ is the amplitude of initial perturbation, $k$ is wave number. We study the KH instability in the following two cases: $\rho _{inner}<\rho_{outer}$ and $\rho _{inner}>\rho _{outer}$.

\subsubsection{Case of $\rho_{inner}<\rho_{outer}$}

In the subsequent simulation, we choose $\rho_{inner}=0.5$, $\rho_{outer}=1.0$, $\mathbf{u}_{inner}=0.5\mathbf{e}_{\theta}$, $\mathbf{u}_{outer}=-0.5\mathbf{e}_{\theta }$, $u_{0}=0.5$,
$D_{\rho}=D_{u}=0.1$, $R_{1}=1$, $R_{2}=2$, $R=1.5$, $k=16$, $\tau =\Delta t=10^{-5}$,
$N_{r}\times N_{\theta}=200\times 90$.

%%%%%%%%%%%%%%%%%%%%%%%%%%%%%%%%%%%%%%%%%%%%%%%%%%%%%%%%%%%%%%%%%%%%
\begin{figure}[tbp]
\center\includegraphics*[bbllx=0pt,bblly=245pt,bburx=595pt,bbury=475pt,angle=0,width=0.99\textwidth]{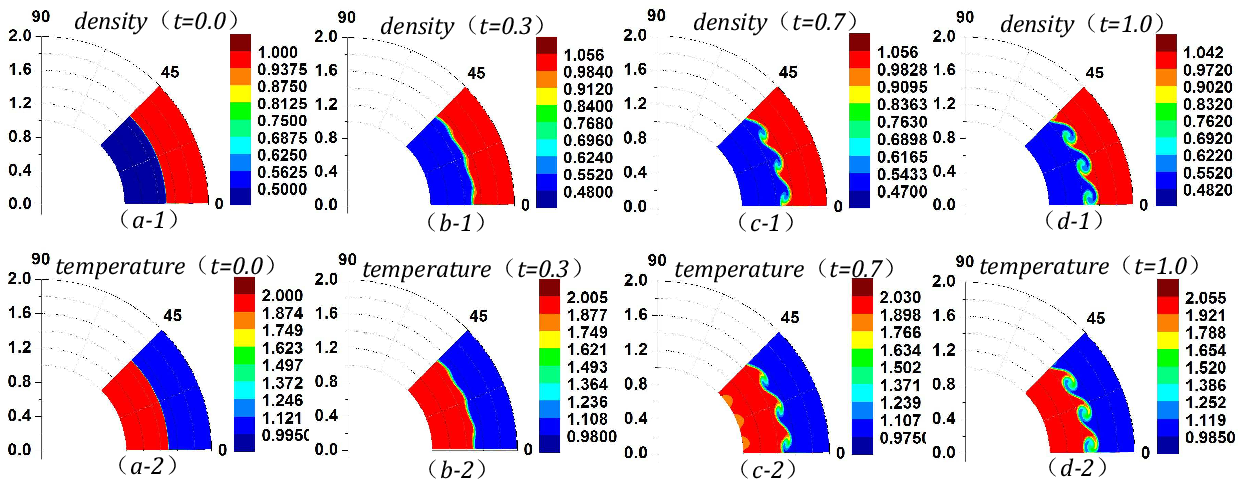}
\caption{Snapshots of KH instability for the case $\protect\rho _{inner}<\protect\rho _{outer}$. The four columns are for the density and temperature contours at $t=0$, $0.3$, $0.7$, and $1$, respectively.}
\label{Fig11}
\end{figure}
\begin{figure}[tbp]
\center\includegraphics*[bbllx=0pt,bblly=0pt,bburx=595pt,bbury=505pt,angle=0,width=0.95\textwidth]{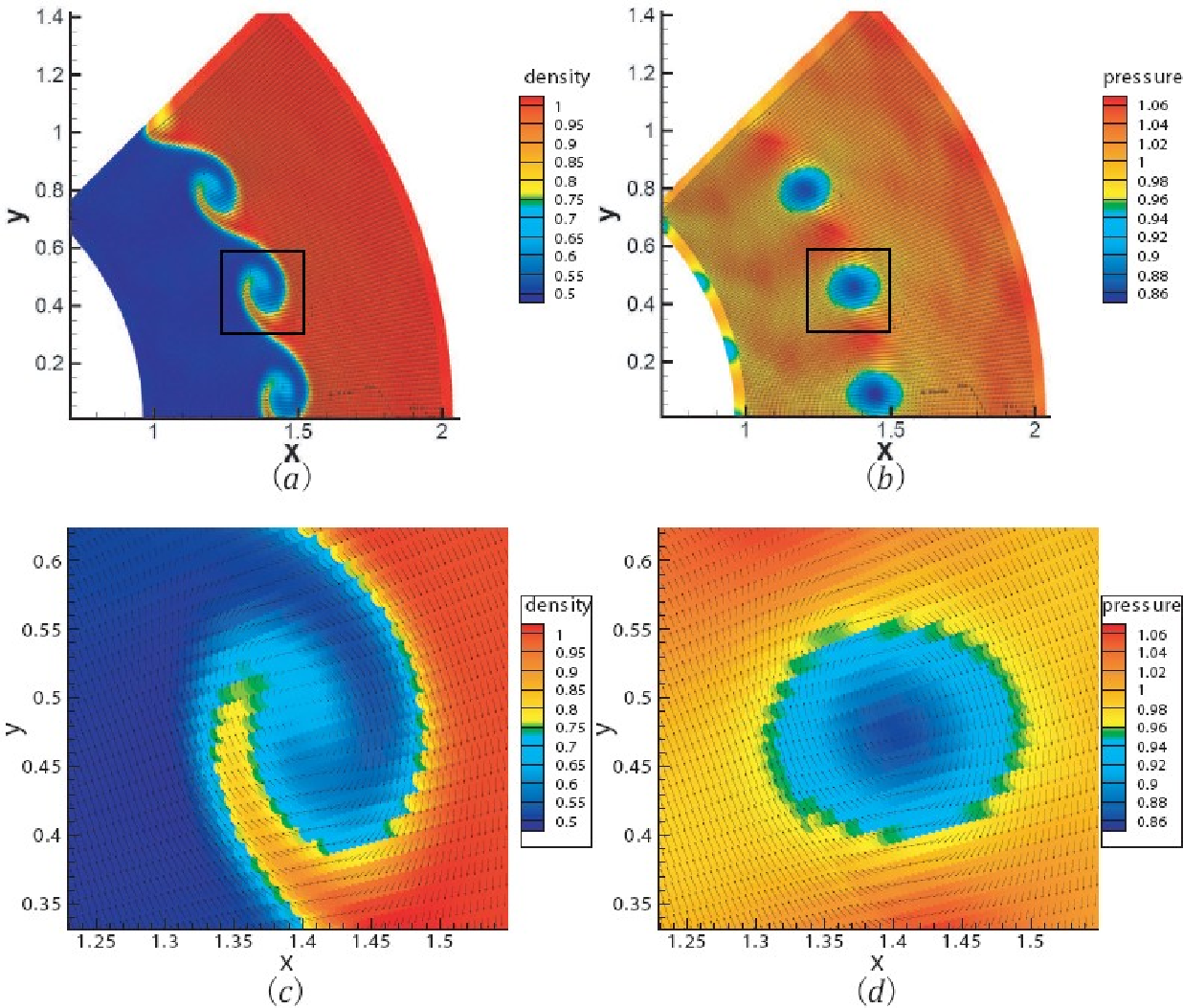}
\caption{Snapshots at time $t=1$ for the case $\rho _{inner}<\rho _{outer}$.
(a) and (b) show the contours of density and pressure in the velocity field, respectively.
(c) shows more clearly the contour of density and velocity field in the region labeled by the square in (a).
(d) shows more clearly the contour of pressure and velocity field in the region labeled by the square in (b).}
\label{Fig12}
\end{figure}
%%%%%%%%%%%%%%%%%%%%%%%%%%%%%%%%%%%%%%%%%%%%%%%%%%%%%%%%%%%%%%%%%%%%
Figure \ref{Fig11} shows the density and temperature contours at the times, $t=0$, $0.3$, $0.7$, and $1$, respectively. Panel (a) shows the initial density and temperature fields. The material interface starts to roll up gradually under the influence of initial velocity disturbance. Panels (b)-(d) show that the interfacial deformation caused by the KH instability becomes more significant with time.

Let's study the physical field at time $t=1$ in Fig.\ref{Fig12}. Figures (a) and (b) show the contour of density and pressure with velocity field, respectively. Figure (c) shows more clearly the contour of density and velocity field in the region labeled by the square in Fig.(a). Figure (d) shows more clearly the contour of pressure and velocity field in the region labeled by the square in Fig.(b). From the velocity field in Fig.(c) we conceive that the KH instability would continue to develop and promote the intermixing and penetrating of the two fluids at the material interface. It's clear to find in Fig.(d) that the minimum value of pressure is at the center of the vortex. In face, it is the pressure gradient that offers the centripetal force required by the rotating flows.

\subsubsection{Case of $\rho_{inner}>\rho_{outer}$}

\begin{figure}[tbp]
\center\includegraphics*[bbllx=0pt,bblly=231pt,bburx=595pt,bbury=476pt,angle=0,width=0.99\textwidth]{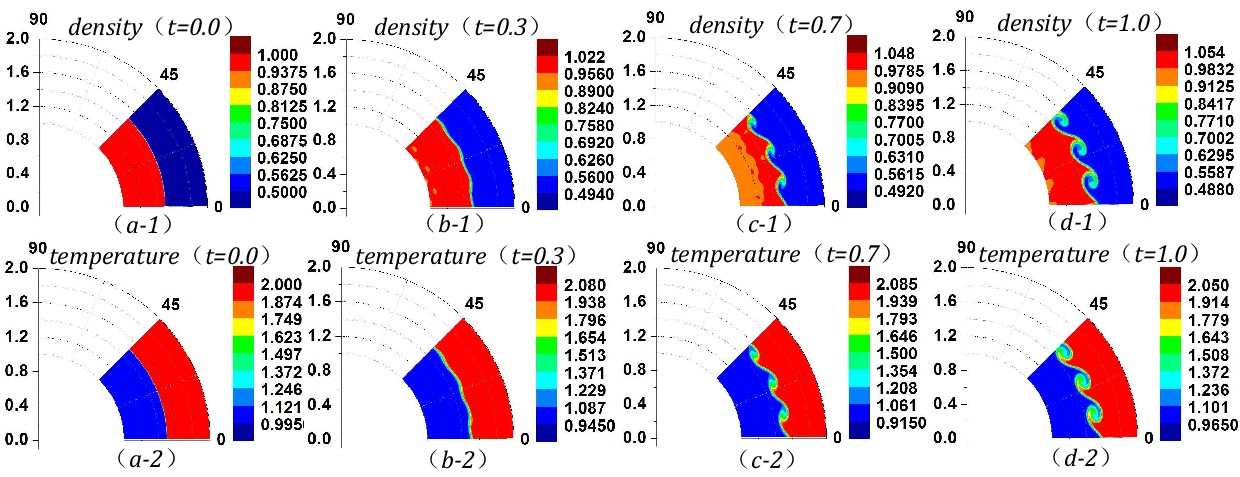}
\caption{Snapshots of KH instability for the case $\rho _{inner}>\rho _{outer}$. The four columns are for the density and temperature contours at $t=0$, $0.3$, $0.7$, and $1$, respectively.}
\label{Fig13}
\end{figure}

In the subsequent simulation, $\rho _{inner}=1.0$, $\rho _{outer}=0.5$, other parameters are the same as those in the case $\rho _{inner}<\rho_{outer}$. Figure \ref{Fig13} shows the contours of density and temperature at $t=0$, $0.3$, $0.7 $ and $1$, respectively. The evolution of KH instability in Fig.\ref{Fig13} is similar to the one in Fig.\ref{Fig11}. From Figs.\ref{Fig11} and \ref{Fig13} we find that the structures within the heavy medium are relatively sharp, likely \textquotedblleft finger"; while the ones within the light medium are relatively smooth, likely \textquotedblleft bubble".
%%%%%%%%%%%%%%%%%%%%%%%%%%%%%%%%%%%%%%%%%%%%%%%%%%%%%%%%%%%%%%%%%%%%
\begin{figure}[tbp]
\center\includegraphics*[bbllx=0pt,bblly=0pt,bburx=595pt,bbury=505pt,angle=0,width=0.95\textwidth]{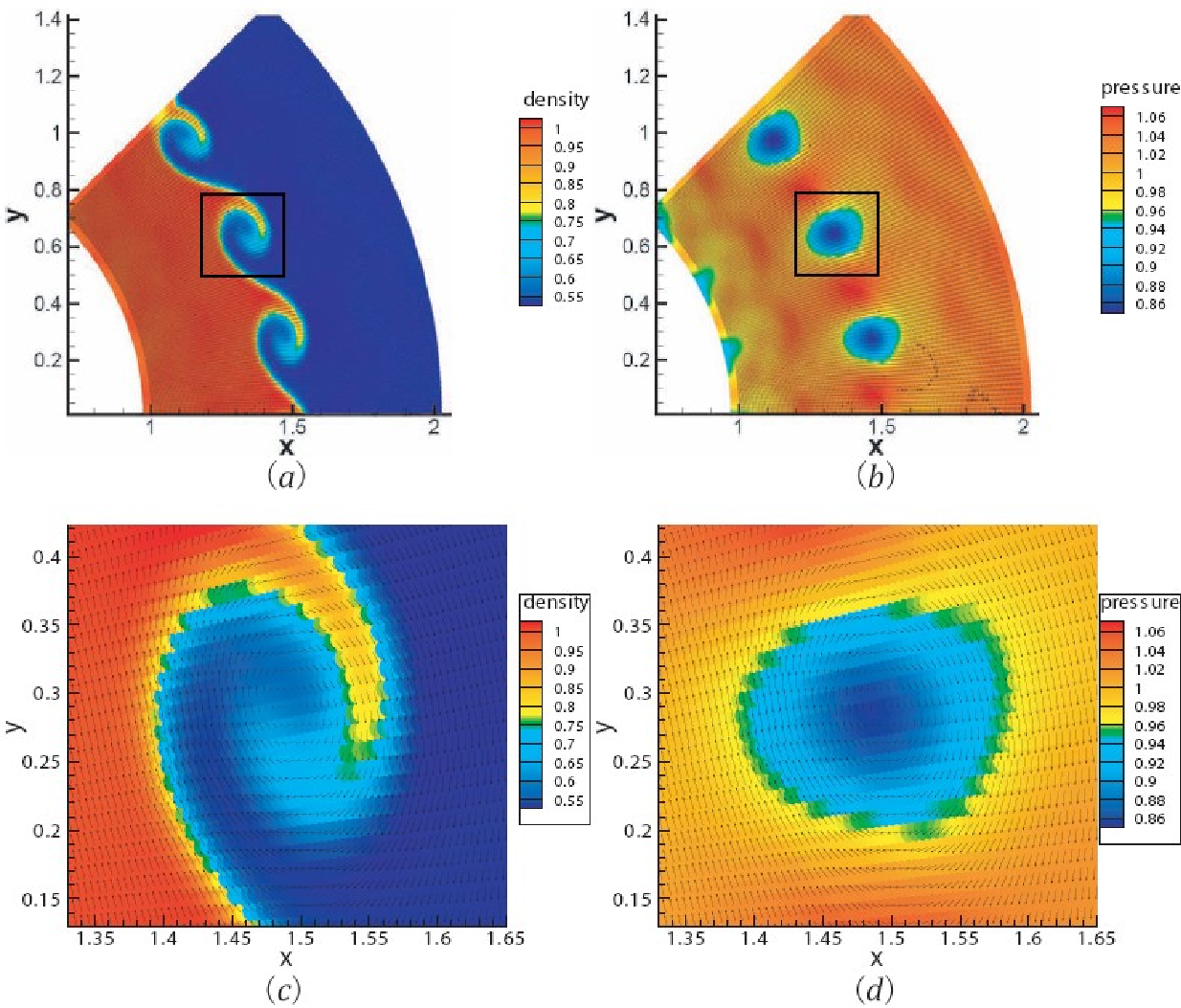}
\caption{Snapshots at time $t=1$ for the case $\rho _{inner}>\rho _{outer}$.
(a) and (b) show the contours of density and pressure in velocity field, respectively.
(c) shows more clearly the contour of density and velocity field in the region labeled by the square in (a).
(d) shows more clearly the contour of pressure and velocity field in the region labeled by the square in (b). }
\label{Fig14}
\end{figure}
%%%%%%%%%%%%%%%%%%%%%%%%%%%%%%%%%%%%%%%%%%%%%%%%%%%%%%%%%%%%%%%%%%%%

Figure \ref{Fig14} shows the contours of density and pressure with velocity field at the time $t=1$. From Figs.\ref{Fig12} and \ref{Fig14} we find that the minimum value of pressure is at the vortex center and its maximum value is at the junction of vortices.

It should be pointed out that in the case $\rho_{inner} > \rho_{outer}$, besides the KH instability, the Rayleigh-Taylor instability also plays a role in the evolution of the rotating flows. Because the material inertia presents an acceleration pointing to the light medium from the heavy one. But since the observation time is short, what we observe is mainly the result of the KH instability.

\section{Non-equilibrium characteristics in two specific cases}

To show the merit of the LB model over traditional numerical models, in this section we study the non-equilibrium characteristics in two specific cases. Among the seven moment relations, Eqs.\eqref{moment1}-\eqref{moment7}, required by our model, only for the first three the equilibrium distribution function $f_{ki}^{eq}$ can be replaced by the distribution function $f_{ki}$. If we replace $f_{ki}^{eq}$ by $f_{ki}$ in the left hand side of any one of Eqs.\eqref{moment4}-\eqref{moment7}, the left and right hand sides of Eqs.\eqref{moment4}-\eqref{moment7} will no longer be in balance. This mismatch measures the departure of the system from local thermodynamic equilibrium.

We define two kinds of space-time dependent fields, moments $\mathbf{M}_{m}$ and central moments $\mathbf{M}_{m}^{\ast}$, as given below:
\begin{eqnarray}
&&\left\{
\begin{array}{l}
\mathbf{M}_{2}(f_{ki})=\sum_{ki}f_{ki}\mathbf{v}_{ki}\mathbf{v}_{ki} \\
\mathbf{M}_{3}(f_{ki})=\sum_{ki}f_{ki}\mathbf{v}_{ki}\mathbf{v}_{ki}\mathbf{v}_{ki} \\
\mathbf{M}_{3,1}(f_{ki})=\sum_{ki}\frac{1}{2}f_{ki}\mathbf{v}_{ki}\cdot \mathbf{v}_{ki}\mathbf{v}_{ki} \\
\mathbf{M}_{4,2}(f_{ki})=\sum_{ki}\frac{1}{2}f_{ki}\mathbf{v}_{ki}\cdot \mathbf{v}_{ki}\mathbf{v}_{ki}\mathbf{v}_{ki}
\end{array}
\right.
\label{M_relative} \\
&&\left\{
\begin{array}{l}
\mathbf{M}_{2}^{\ast }(f_{ki})=\sum_{ki}f_{ki}(\mathbf{v}_{ki}-\mathbf{u})(\mathbf{v}_{ki}-\mathbf{u}) \\
\mathbf{M}_{3}^{\ast}(f_{ki})=\sum_{ki}f_{ki}(\mathbf{v}_{ki}-\mathbf{u})(\mathbf{v}_{ki}-\mathbf{u})(\mathbf{v}_{ki}-\mathbf{u}) \\
\mathbf{M}_{3,1}^{\ast }(f_{ki})=\sum_{ki}\frac{1}{2}f_{ki}(\mathbf{v}_{ki}-\mathbf{u})\cdot (\mathbf{v}_{ki}-\mathbf{u})(\mathbf{v}_{ki}-\mathbf{u)} \\
\mathbf{M}_{4,2}^{\ast }(f_{ki})=\sum_{ki}\frac{1}{2}f_{ki}(\mathbf{v}_{ki}-\mathbf{u})\cdot (\mathbf{v}_{ki}-\mathbf{u})(\mathbf{v}_{ki}-\mathbf{u)}(\mathbf{v}_{ki}-\mathbf{u)}
\end{array}
\right.
\label{M_absolute}
\end{eqnarray}
where the subscript $``3,1"$ means that the $3$rd-order tensor is contracted to a $1$st-order tensor and the similar is for $``4,2"$. The moment $\mathbf{M}_{3,1}$($=M_{3,1,\alpha} \mathbf{e}_{\alpha}$) is a vector. It has two components, $M_{3,1,r}$ and $M_{3,1,\theta}$. The moment $\mathbf{M}_{2}$($=M_{2,\alpha \beta}\mathbf{e}_{\alpha }\mathbf{e}_{\beta }$) is a second-order tensor with four components. Among the four components, only three, $M_{2,rr}$, $M_{2,r\theta}$ and $M_{2,\theta\theta}$, are independent.
The case for the moment $\mathbf{M}_{4,2}$($=M_{4,2,\alpha \beta}\mathbf{e}_{\alpha}\mathbf{e}_{\beta}$) is similar. The moment $\mathbf{M}_{3}$($=M_{3,\alpha \beta \gamma}\mathbf{e}_{\alpha}\mathbf{e}_{\beta}\mathbf{e}_{\gamma}$) is a third-order tensor with eight components. Among the eight components, only four, $M_{3,rrr}$, $M_{3,rr\theta}$, $M_{3,r\theta\theta}$ and $M_{3,\theta\theta\theta}$, are independent. The central moments $\mathbf{M}_{m}^{\ast}$ are mathematically similar to $\mathbf{M}_{m}$.

In probability theory, for the one-dimensional distribution function $f(v)$, the central moment $M_{3}^{\ast}=\int dv f(v) (v-u)^3$ is called ``skewness". The fourth-order central moment $M_{4}^{\ast}=\int dv f(v) (v-u)^4$ describes the ``flatness" of the distribution and is called ``kurtosis". For a Gaussian distribution function, $f(v)= 1/\sqrt{2\pi} \exp[-(v-u)^{2}/2]$, $M_{4}^{\ast}=3$.
For the case with  $M_{4}^{\ast}>3$ and  $M_{2}^{\ast}=1$, the distribution is sharper than the Gaussian at the central position.

Physically, all moments above associates with the variance of the distribution function.
The trace of moment $\mathbf{M}_{2}$ associates with temperature and its off-diagonal
components associate with the shear effects. The former is a conserved quantity.
When the system is not in its thermodynamic equilibrium state, the latter may not be zero.
The similar is for central moment $\mathbf{M}_{2}^{\ast}$.
The moment $\mathbf{M}_{3}$ associates with the heat flux resulting from macroscopic flow
and ``energy flow caused by microscopic fluctuation". For an equilibrium state, it only
describes convection of energy resulting from macroscopic behavior. For the non-equilibrium state,
besides that energy convection, it also includes ``energy flow caused by microscopic fluctuation".
The central moment $\mathbf{M}_{3}^{\ast}$ only describes ``energy flow caused by microscopic fluctuation".
Therefore $\mathbf{M}_{3}^{\ast}=0$ in an equilibrium state.
The moment $\mathbf{M}_{3,1}$ and central moment $\mathbf{M}_{3,1}^{\ast}$ is a contraction
of $\mathbf{M}_{3}$ and $\mathbf{M}_{3}^{\ast}$, respectively.
For the central moments $\mathbf{M}_{3}^{\ast}$ and $\mathbf{M}_{3,1}^{\ast}$, a breaking of
the $f(\mathbf{v}) = f(\mathbf{-v})$ symmetry allows to eventually transport heat without
necessarily carrying a net flow. In addition, the third-order central moment $\mathbf{M}_{3}^{\ast}$ may not be zero,
while the first-order central moment $\mathbf{M}_{1}^{\ast}=\sum_{ki}f_{ki}(\mathbf{v}_{ki}-\mathbf{u})$ must be zero.

By Galilean invariance, it is clear that the moment $\mathbf{M}_{m}$ contains the information of
macroscopic flow velocity $\mathbf{u}$, while the moment $\mathbf{M}_{m}^{\ast }$ is only the
manifestation of the thermo-fluctuations of molecules relative to the macroscopic flow velocity $\mathbf{u}$.

The manifestations of deviating from thermodynamic equilibrium from the two kinds of moments are as below:
\begin{eqnarray}
\mathbf{\Delta }_{m}&=&\mathbf{M}_{m}(f_{ki})-\mathbf{M}_{m}(f_{ki}^{eq})=
\mathbf{M}_{m}(f_{ki}-f_{ki}^{eq})
\label{D_relative} \\
\mathbf{\Delta }_{m}^{\ast }&=&\mathbf{M}_{m}^{\ast }(f_{ki})-\mathbf{M}_{m}^{\ast }(f_{ki}^{eq})=
\mathbf{M}_{m}^{\ast }(f_{ki}-f_{ki}^{eq})
\label{D_absolute}
\end{eqnarray}
Similarly, $\mathbf{\Delta }_{m}$ contains the information of the macroscopic flow velocity $\mathbf{u}$,
while $\mathbf{\Delta }_{m}^{\ast }$ does not.

\subsection{Simulation results and analysis}

\begin{figure}[tbp]\center\includegraphics*[bbllx=0pt,bblly=0pt,bburx=595pt,bbury=258pt,angle=0,width=0.75\textwidth]{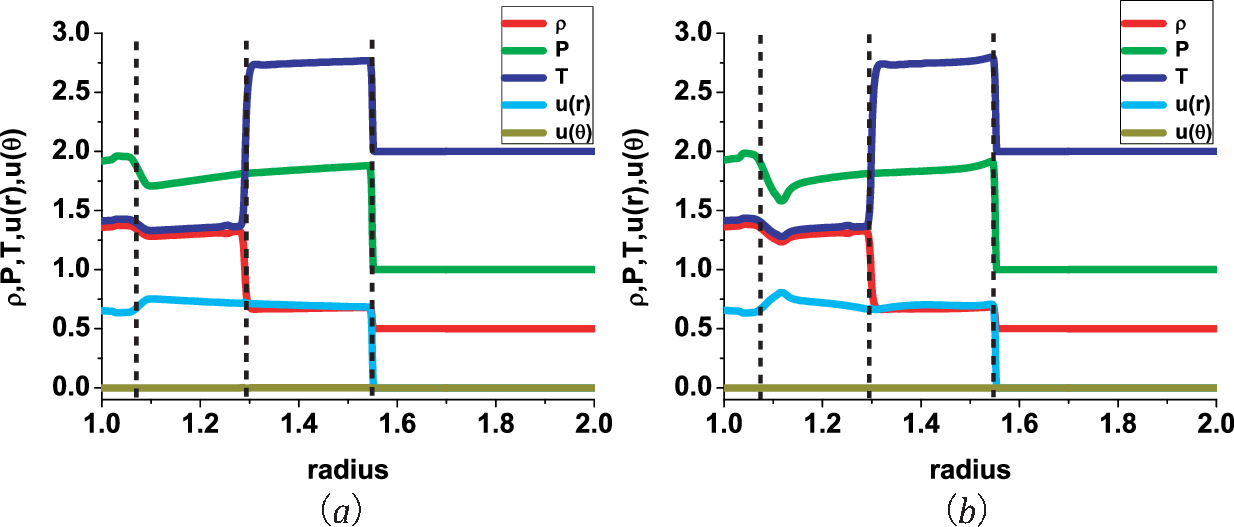}
\caption{Profiles of physical quantities ($\rho $, $P$, $T$, $u_{r}$, $u_{\theta }$) in the case of the shock wave travelling outwards from the heavy medium to the light one at the time $t=0.15$. (a) Without initial perturbation at the material interface. (b) With initial sinusoidal perturbation at the material interface. Three lines are shown to guide the eyes for the three interfaces.}
\label{Fig15}
\end{figure}

Now, we study the dynamic procedure where a shock wave propagates outwards from the heavy material to the light one. As the first step, we study the simplest situation where the incident shock wave is perpendicular to the unperturbed circular interface. In the second case, the interface is perturbed sinusoidally, and consequently the RM instability will occur. We choose such a time, $t=0.15$, when the system shows three different interfaces, see Fig.\ref{Fig15}. From left to right, the first is for the rarefaction wave, the second is for the material interface, the third is for the shock wave.
%%%%%%%%%%%%%%%%%%%%%%%%%%%%%%%%%%%%%%%%%%%%%%%%%%%%%%%%%%%%%%%%%%%%
%%%%%%%%%%%%%%%%%%%%%%%%%%%%%%%%%%%%%%%%%%%%%%%%%%%%%%%%%%%%%%%%%%%%
\begin{figure}[tbp]
\center\includegraphics*[bbllx=0pt,bblly=0pt,bburx=595pt,bbury=282pt,angle=0,width=0.99\textwidth]{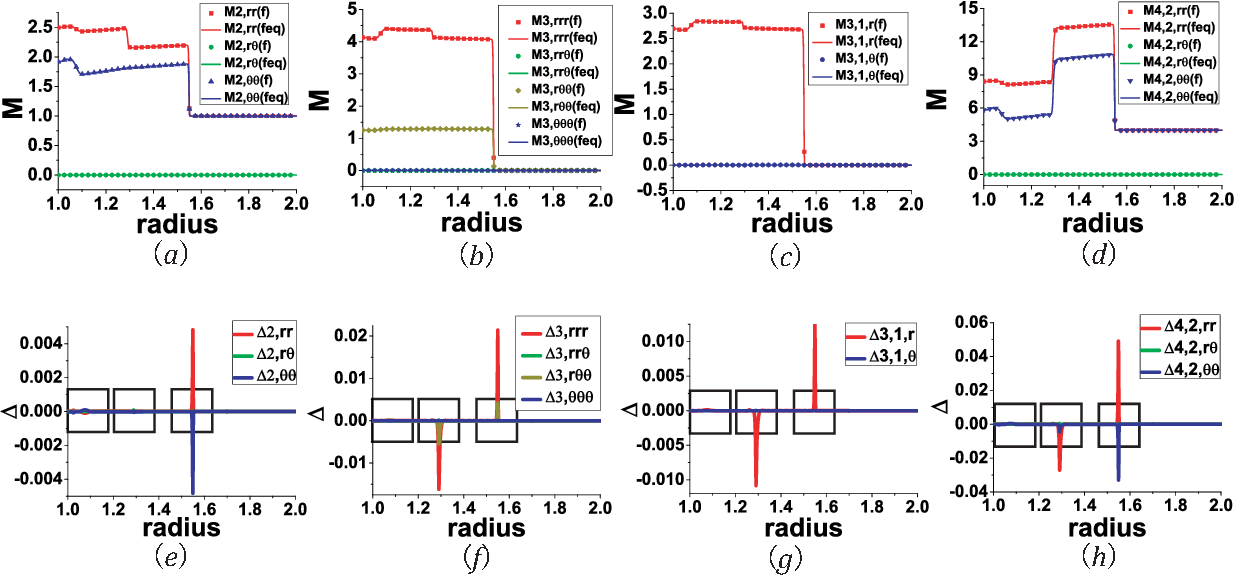}
\caption{Moments and their corresponding non-equilibrium manifestations for the case without initial perturbation at the material interface. The time $t=0.15$. Figures (a)-(d) are for $\mathbf{M}_{2}$, $\mathbf{M}_{3}$, $\mathbf{M}_{3,1}$, $\mathbf{M}_{4,2}$, respectively. The symbols are for moments from $f_{ki}$ and the solid lines are for moments from $f_{ki}^{eq}$. Figures (e)-(h) are
for deviations $\mathbf{\Delta }_{2}$, $\mathbf{\Delta }_{3}$, $\mathbf{\Delta }_{3,1}$, $\mathbf{\Delta }_{4,2}$, respectively. Only independent components of $\mathbf{M}_{m}$ and $\mathbf{\Delta } _{m}$ are shown. The specific correspondences are referred to the legends. Three squares are shown to guide the eyes for the interfaces.}
\label{Fig16}
\end{figure}
%%%%%%%%%%%%%%%%%%%%%%%%%%%%%%%%%%%%%%%%%%%%%%%%%%%%%%%%%%%%%%%%%%%%
\begin{figure}[tbp]
\center\includegraphics*[bbllx=0pt,bblly=0pt,bburx=595pt,bbury=385pt,angle=0,width=0.99\textwidth]{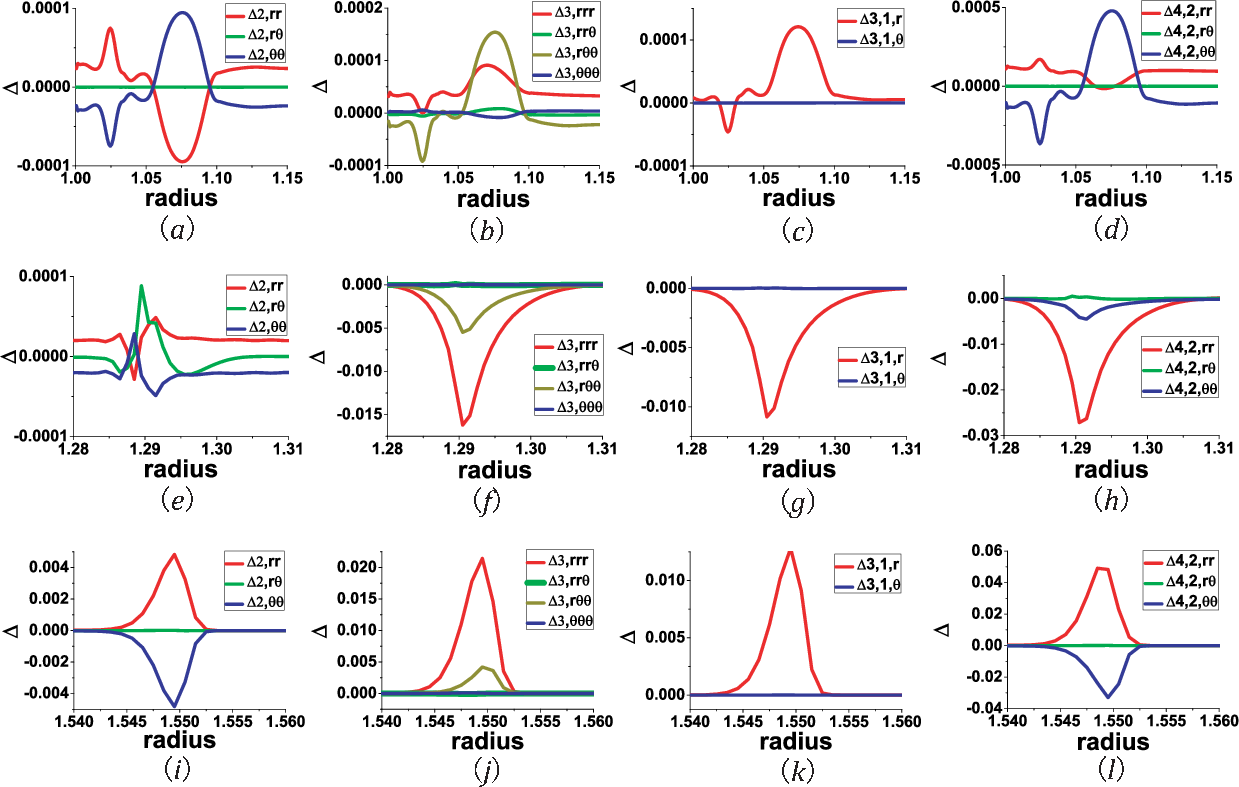}
\caption{The deviations $\mathbf{\Delta}_m$ versus the radius, which are enlargements
of the portions labeled by squares in Figs.\ref{Fig16}(e)-(h).
Figures (a)-(d) are for the region around the first interface, with $1.00\leq r\leq 1.15$.
Figures (e)-(h) are for the region around the second interface, with $1.28\leq r\leq1.31$.
Figures (i)-(l) are for the region around the third interface, with $1.54\leq r\leq 1.56$.}
\label{Fig17}
\end{figure}
%%%%%%%%%%%%%%%%%%%%%%%%%%%%%%%%%%%%%%%%%%%%%%%%%%%%%%%%%%%%%%%%%%%%
\begin{figure}[tbp]
\center\includegraphics*[bbllx=0pt,bblly=0pt,bburx=595pt,bbury=280pt,angle=0,width=0.99\textwidth]{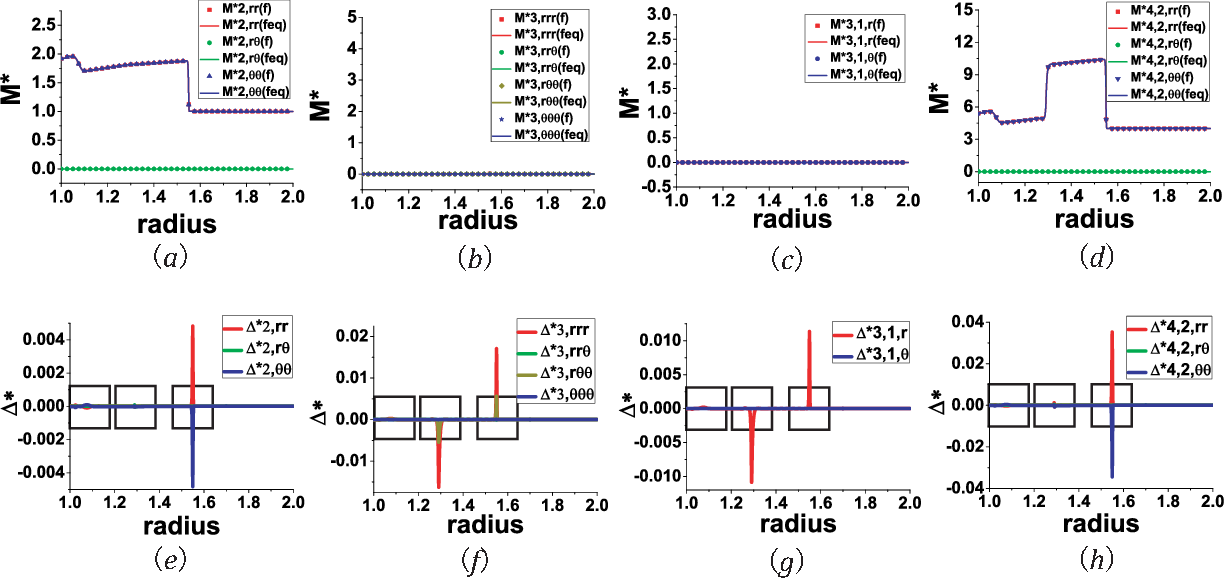}
\caption{Central moments and their corresponding non-equilibrium manifestations for the case without initial perturbation at the material interface. The time $t=0.15$. Figures (a)-(d) are for $\mathbf{M}_{2}^{\ast}$, $\mathbf{M}_{3}^{\ast}$, $\mathbf{M}_{3,1}^{\ast}$, $\mathbf{M}_{4,2}^{\ast}$, respectively. The symbols are for central moments from $f_{ki}$ and the solid lines are for central moments from $f_{ki}^{eq}$. Figures (e)-(h) are for deviations $\mathbf{\Delta}_{2}^{\ast}$, $\mathbf{\Delta }_{3}^{\ast}$, $\mathbf{\Delta}_{3,1}^{\ast}$, $\mathbf{\Delta}_{4,2}^{\ast}$, respectively. Only independent components of $\mathbf{M}_{m}^{\ast}$ and $\mathbf{\Delta}_{m}^{\ast}$ are shown. The specific correspondences are referred to the legends. Three squares are shown to guide the eyes for the interfaces.}
\label{Fig18}
\end{figure}
%%%%%%%%%%%%%%%%%%%%%%%%%%%%%%%%%%%%%%%%%%%%%%%%%%%%%%%%%%%%%%%%%%%%
\begin{figure}[tbp]
\center\includegraphics*[bbllx=0pt,bblly=0pt,bburx=595pt,bbury=385pt,angle=0,width=0.99\textwidth]{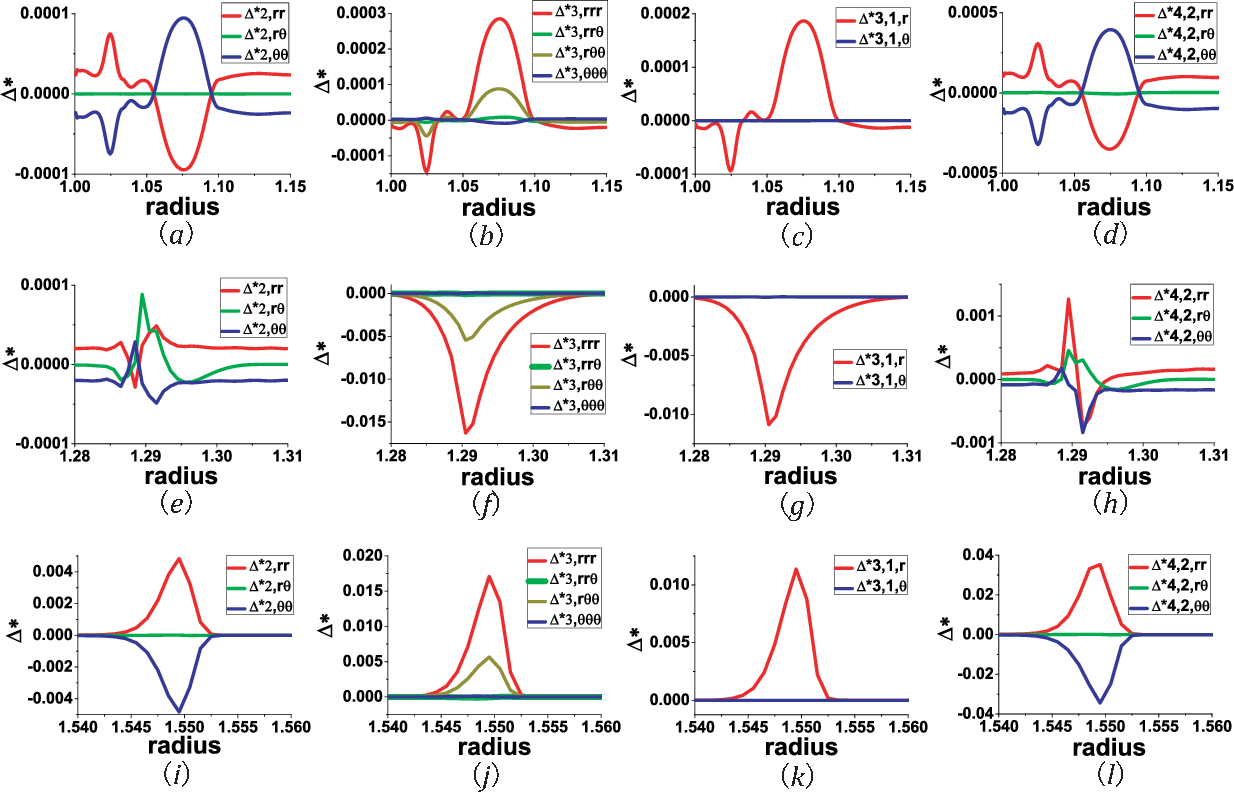}
\caption{The deviations $\mathbf{\Delta}^*_m$ versus the radius, which are enlargements
of the portions labeled by squares in Figs.\ref{Fig18}(e)-(h).
Figures (a)-(d) are for the region around the first interface, with $1.00\leq r\leq 1.15$.
Figures (e)-(h) are for the region around the second interface, with $1.28\leq r\leq1.31$.
Figures (i)-(l) are for the region around the third interface, with $1.54\leq r\leq 1.56$.}
\label{Fig19}
\end{figure}
%%%%%%%%%%%%%%%%%%%%%%%%%%%%%%%%%%%%%%%%%%%%%%%%%%%%%%%%%%%%%%%%%%%%
\begin{figure}[tbp]
\center\includegraphics*[bbllx=0pt,bblly=0pt,bburx=595pt,bbury=282pt,angle=0,width=0.99\textwidth]{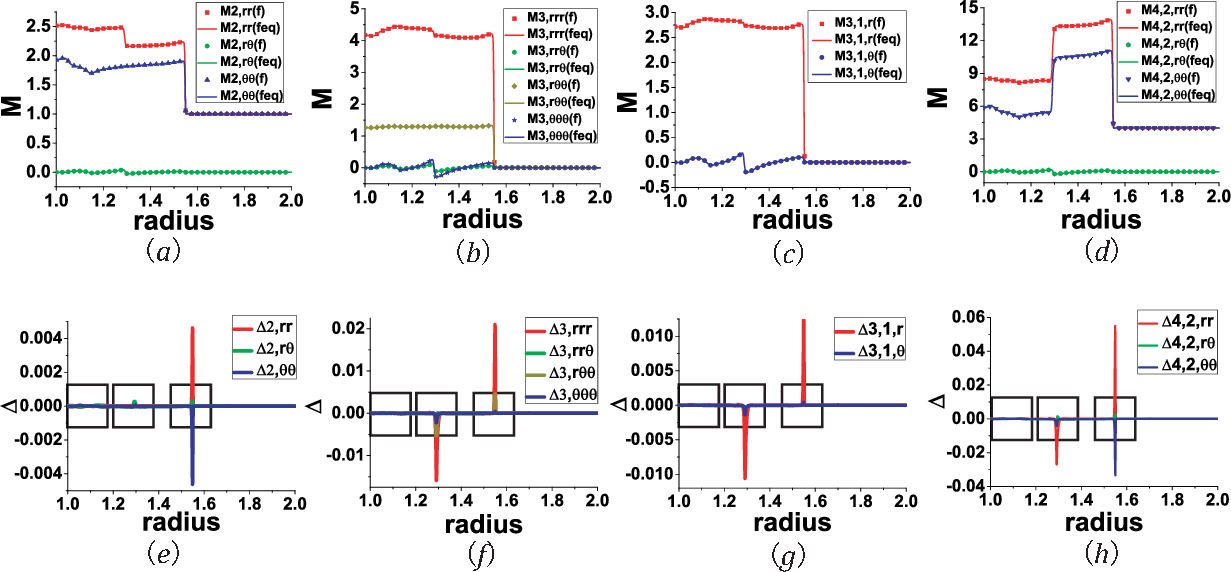}
\caption{Moments and their corresponding non-equilibrium manifestations for the case with initial sinusoidal perturbation at the material interface. The time $t=0.15$. Figures (a)-(d) are for $\mathbf{M}_{2}$, $\mathbf{M}_{3}$, $\mathbf{M}_{3,1}$, $\mathbf{M}_{4,2}$, respectively. The symbols are for moments from $f_{ki}$ and the solid lines are for moments from $f_{ki}^{eq}$.
Figures (e)-(h) are for deviations $\mathbf{\Delta}_{2}$, $\mathbf{\Delta}_{3}$, $\mathbf{\Delta}_{3,1}$, $\mathbf{\Delta }_{4,2}$, respectively. Only independent components of $\mathbf{M}_{m}$ and $\mathbf{\Delta }_{m}$ are shown. The specific correspondences are referred to the legends. Three squares are shown to guide the eyes for the interfaces.}
\label{Fig20}
\end{figure}
%%%%%%%%%%%%%%%%%%%%%%%%%%%%%%%%%%%%%%%%%%%%%%%%%%%%%%%%%%%%%%%%%%%%
\begin{figure}[tbp]
\center\includegraphics*[bbllx=0pt,bblly=0pt,bburx=595pt,bbury=385pt,angle=0,width=0.99\textwidth]{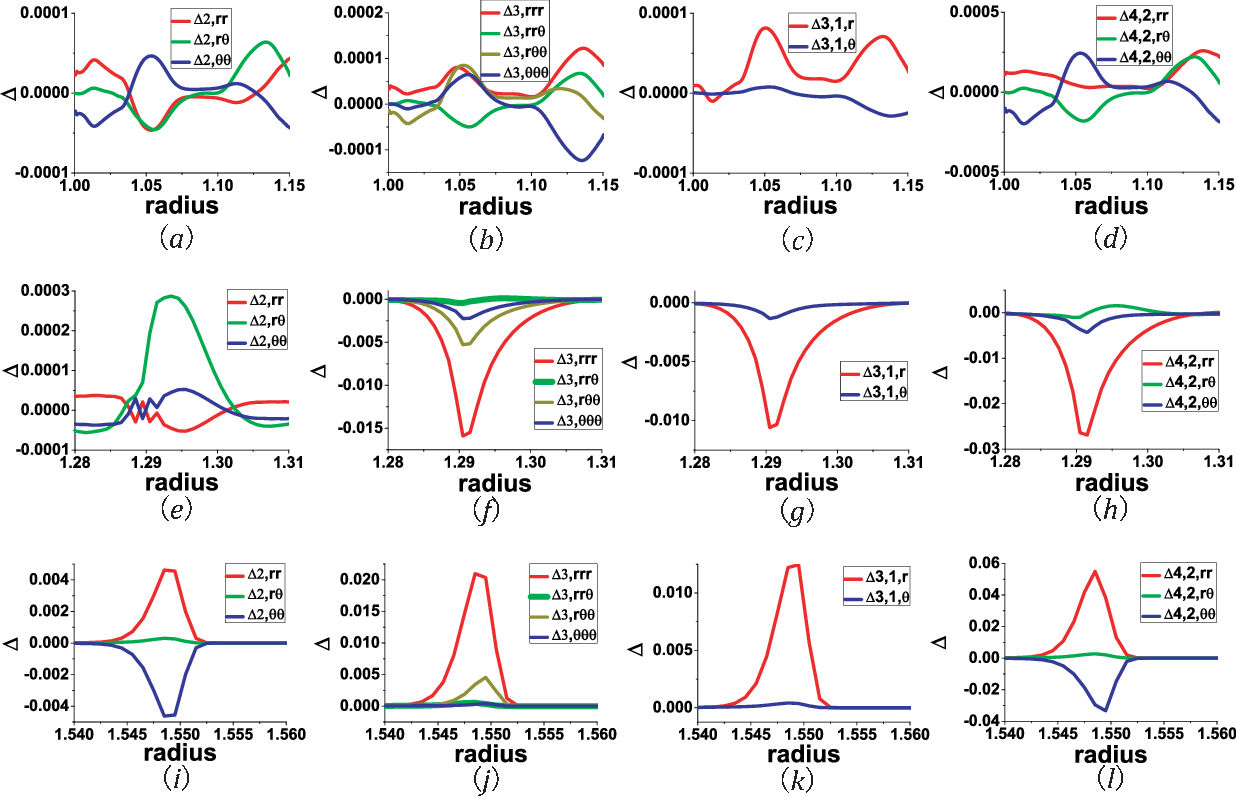}
\caption{The deviations $\mathbf{\Delta}_m$ versus the radius, which are enlargements
of the portions labeled by squares in Figs.\ref{Fig20}(e)-(h).
Figures (a)-(d) are for the region around the first interface, with $1.00\leq r\leq 1.15$.
Figures (e)-(h) are for the region around the second interface, with $1.28\leq r\leq1.31$.
Figures (i)-(l) are for the region around the third interface, with $1.54\leq r\leq 1.56$.}
\label{Fig21}
\end{figure}
%%%%%%%%%%%%%%%%%%%%%%%%%%%%%%%%%%%%%%%%%%%%%%%%%%%%%%%%%%%%%%%%%%%%
\begin{figure}[tbp]
\center\includegraphics*[bbllx=0pt,bblly=0pt,bburx=595pt,bbury=285pt,angle=0,width=0.99\textwidth]{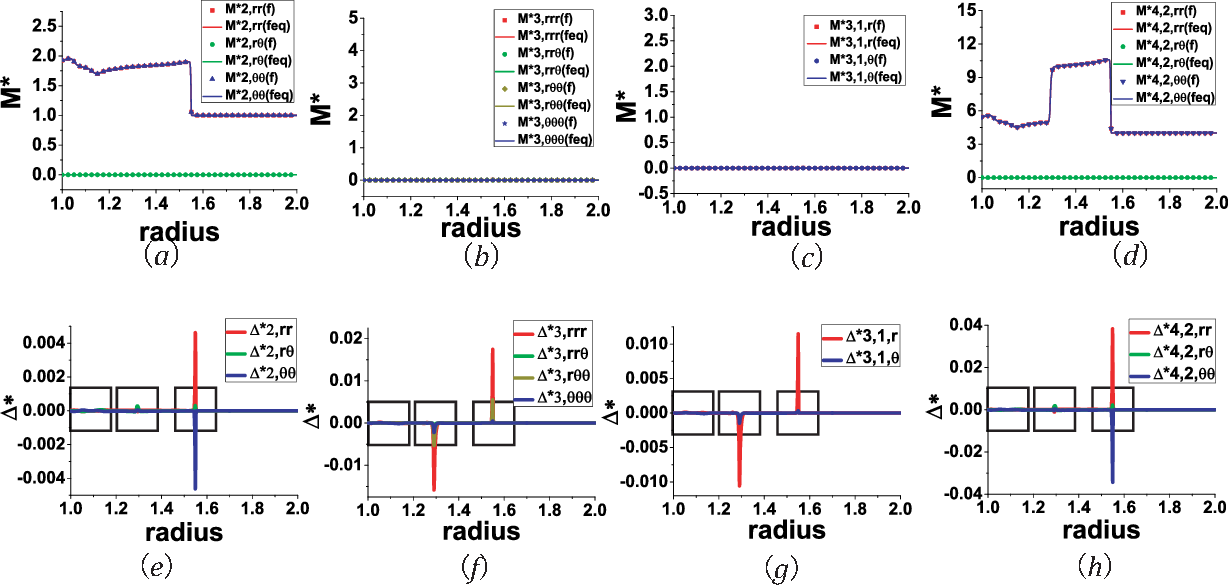}
\caption{Central moments and their corresponding non-equilibrium manifestations for the case with initial sinusoidal perturbation at the material interface. The time $t=0.15$. Figures (a)-(d) are for $\mathbf{M}_{2}^{\ast}$, $\mathbf{M}_{3}^{\ast}$, $\mathbf{M}_{3,1}^{\ast}$, $\mathbf{M}_{4,2}^{\ast}$, respectively. The symbols are for central moments from $f_{ki}$ and the solid lines are for central moments from $f_{ki}^{eq}$. Figures (e)-(h) are for deviations $\mathbf{\Delta }_{2}^{\ast}$,
$\mathbf{\Delta }_{3}^{\ast}$, $\mathbf{\Delta}_{3,1}^{\ast}$, $\mathbf{\Delta }_{4,2}^{\ast}$, respectively. Only independent components of $\mathbf{M}_{m}^{\ast}$ and $\mathbf{\Delta } _{m}^{\ast}$ are shown. The specific correspondences are referred to the legends. Three squares are shown to guide the eyes for the interfaces.}
\label{Fig22}
\end{figure}
%%%%%%%%%%%%%%%%%%%%%%%%%%%%%%%%%%%%%%%%%%%%%%%%%%%%%%%%%%%%%%%%%%%%
\begin{figure}[tbp]
\center\includegraphics*[bbllx=0pt,bblly=0pt,bburx=595pt,bbury=385pt,angle=0,width=0.99\textwidth]{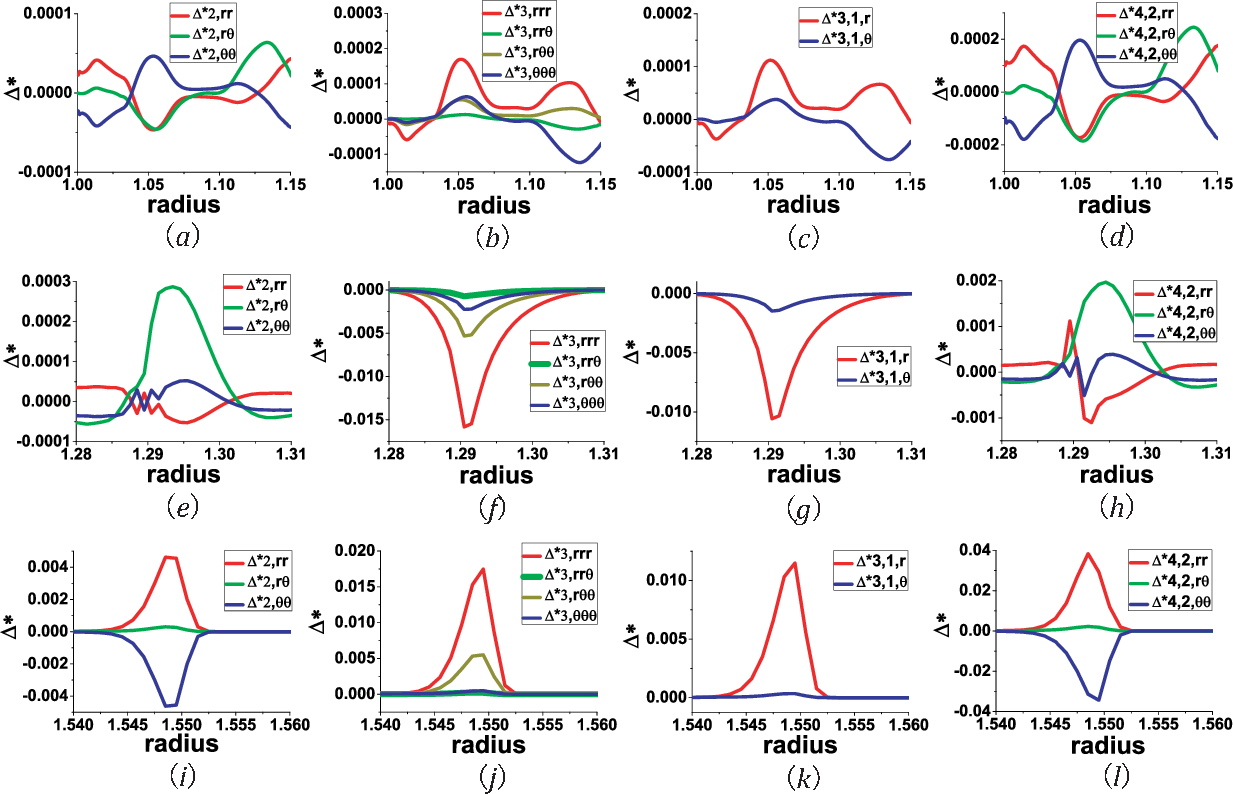}
\caption{The deviations $\mathbf{\Delta}^*_m$ versus the radius, which are enlargements
of the portions labeled by squares in Figs.\ref{Fig22}(e)-(h).
Figures (a)-(d) are for the region around the first interface, with $1.00\leq r\leq 1.15$.
Figures (e)-(h) are for the region around the second interface, with $1.28\leq r\leq1.31$.
Figures (i)-(l) are for the region around the third interface, with $1.54\leq r\leq 1.56$.}
\label{Fig23}
\end{figure}
%%%%%%%%%%%%%%%%%%%%%%%%%%%%%%%%%%%%%%%%%%%%%%%%%%%%%%%%%%%%%%%%%%%%

Figure \ref{Fig15} shows the profiles of physical quantities ($\rho $, $P$, $T$, $u_{r}$, $u_{\theta }$) along the radius with the fixed azimuthal angle $\theta =7\pi/48$. At this azimuthal angle the perturbation amplitude is close to zero. Figure (a) is for the case without initial perturbation at the material interface. Figure (b) is for the case with initial sinusoidal perturbation at the material interface. Three lines are shown to guide the eyes for the three interfaces. From Fig.\ref{Fig15} we can find the steep variations of physical quantities at the three interfaces. For the case without perturbation at the material interface, we show
the results of $\mathbf{M}_{m}$ and $\mathbf{\Delta} _{m}$ in Fig.\ref{Fig16}. All independent components of $\mathbf{M}_{m}$, and $\mathbf{\Delta }_{m}$ are shown. The specific correspondences are referred to the legends.
The 12 plots in Fig.\ref{Fig17} are the enlargements of the 12 portions labeled by the 12 squares in Fig.\ref{Fig16}.
Figures \ref{Fig17}(a)-(d) correspond to the portions labeled by the first squares in Figs.\ref{Fig16}(e)-(h), respectively.
Figures \ref{Fig17}(e)-(h) correspond to the portions labeled by the second squares in Figs.\ref{Fig16}(e)-(h), respectively.
Figures \ref{Fig17}(i)-(l) correspond to the portions labeled by the third squares in Figs.\ref{Fig16}(e)-(h), respectively.
The results of $\mathbf{M}_{m}^{\mathbf{\ast}}$ and $\Delta_{m}^{\ast}$  are shown in Fig.\ref{Fig18}.
The 12 plots in Fig.\ref{Fig19} are the enlargements of the 12 portions labeled by the 12 squares in Fig.\ref{Fig18}.
The specific correspondences between Figs.\ref{Fig19} and \ref{Fig18} are similar to the case of Figs.\ref{Fig17} and \ref{Fig16}.
For the case with sinusoidal perturbation at the material interface, along the same radius,
the results of $\mathbf{M}_{m}$ and $\Delta _{m}$ are shown in Fig.\ref{Fig20} and Fig.\ref{Fig21}.
The results of $\mathbf{M}_{m}^{\mathbf{\ast }}$ and $\Delta_{m}^{\ast }$ are shown in Fig.\ref{Fig22} and Fig.\ref{Fig23}.
The specific correspondences between Figs.\ref{Fig21} and \ref{Fig20} and the specific correspondences between Figs.\ref{Fig23} and \ref{Fig22} are also similar to the case of Figs.\ref{Fig17} and \ref{Fig16}.

For both the two cases, one can clearly find the existence of the three interfaces via typical variations of the moments and corresponding moment differences.

(1) Around the shock front, the system starts to deviate from thermodynamic equilibrium once the physical quantities ($\rho$,$T$,$P$,$u$) start to increase, and goes back to its thermodynamic equilibrium as the physical quantities attain their steady values required by the Hugoniot relations. The shocking procedure is very fast and the shock interface is very thin. The changing rates of macroscopic quantities are quite high. Hence, there is little time for the thermo-diffusion process around the shock front and there is little time for the system to relax to its thermodynamic equilibrium. During the shocking precess, $\Delta_{2,rr}$ (or $\Delta_{2,rr}^{\ast}$) shows a positive peak, while $\Delta_{2,\theta \theta}$ (or $\Delta_{2,\theta \theta}^{\ast}$) shows a negative peak with the same amplitude. Meanwhile, $\Delta_{2,r \theta}$ (or $\Delta_{2,r \theta}^{\ast}$) is close to zero.

(2) Around the rarefaction front, the mechanical effect instead of the thermo-diffusion takes a leading role, which is similar to the shock front. In this sense, the two fronts could be named mechanical interfaces. Compared with the shock front, the rarefaction front is much wider and the gradients of physical quantities are smaller. There is more relaxation time for the rarefaction front. Therefore, the system is closer to its thermodynamic equilibrium around the rarefaction front than around the shock front.

(3) Around the material interface, the peak value of $\Delta_{2,rr}$ (or $\Delta_{2,rr}^{\ast}$) is much smaller than the value at the shock front or rarefaction front. Physically, in contrast to the shock or rarefaction procedure, there is enough relaxation time in the process of the thermo-diffusion around the material interface. And the material interface becomes wider and wider.

Further more, $\mathbf{M}_{4,2}$ and $\mathbf{\Delta} _{4,2}$ ($\mathbf{M}_{4,2}^{\mathbf{\ast }}$ and $\Delta_{4,2}^{\ast }$) show similar behavior with $\mathbf{M}_{2}$ and $\mathbf{\Delta}_{2}$ ($\mathbf{M}_{2}^{\ast}$ and $\mathbf{\Delta} _{2}^{\ast }$). Results of $\mathbf{\Delta}_{3}^{\ast }$ and $\mathbf{\Delta }_{3,1}^{\ast }$ can be analyzed in a similar way. The components of $\mathbf{\Delta}_{m}^{\ast }$ can be labeled by $r^{p}\theta ^{q}$, where $p$,$q=1$,$2$,or $3$. At the shock or rarefaction interface, if $q=0$, the corresponding component is the largest. If $q=1$ or $3$, the corresponding component is negligibly small.

Comparing shock front with the rarefaction front in Figs.\ref{Fig16}-\ref{Fig23}, we can find that the shock wave increases density, pressure and temperature, while the rarefaction wave decreases those quantities. In other words, the two waves have opposite mechanical effects. Although around both the two interfaces, from left to right, the values of density, temperature and pressure become smaller, the non-equilibrium manifestations are oppositely different. The physical reason is as follows. The shock wave propagates outwards, while the rarefaction wave propagates inwards. Along their propagation directions, the physical quantities decrease around the shock wave, while they increase around the rarefaction wave.

Compared to the case without initial perturbation in Figs.\ref{Fig16}-\ref{Fig19}, the case with perturbation in Figs.\ref{Fig20}-\ref{Fig23} is much more complex around the material interface. Specially, comparing Fig.\ref{Fig17}(e) and Fig.\ref{Fig21}(e) gives that $\Delta_{2,r\theta}$ has a larger peak value in the latter case. Similarly, the peak value of $\Delta^*_{2,r\theta}$ in Fig.\ref{Fig23}(e) is larger than the one in Fig.\ref{Fig19}(e).

Physically, the initial perturbation enhances the shear viscosity effects in the evolution of RM instability. Hence, the $v_{r}$-$v_{\theta}$ coupling effect is pronounced at the material interface with RM instability.
Other plots in Fig.\ref{Fig17} and in Fig.\ref{Fig21} show consistent information. The information from $\mathbf{\Delta} _{m}$ in Fig.\ref{Fig17} (Fig.\ref{Fig21}) and that from $\mathbf{\Delta}_{m}^{\ast }$ Fig.\ref{Fig19} (Fig.\ref{Fig23}) are complementary.

Via comparing the material interface with the two mechanical interfaces in Figs.\ref{Fig20}-\ref{Fig23}, it's easy to find that
the $v_{r}$-$v_{\theta}$ coupling effect is much more pronounced around the former interface than around the latter two. Physically, there is no tangential motion of flow at the two mechanical interfaces, while there is shearing motion around the material interface.

It should be pointed out that, the situation of the material interface with initial perturbation varies with the azimuthal angle $\theta$. The analysis for other $\theta$ is beyond this work.

All the non-equilibrium effects in Figs.\ref{Fig16}-\ref{Fig23} can be consistently interpreted as follows. Among the four physical fields of density, momentum, pressure and temperature, the gradient of anyone can trigger the non-equilibrium effects. In fact, those gradients seldom appear alone. They will affect each other and couple together to play a role in triggering non-equilibrium. Here we give an explanation of the non-equilibrium effects by the temperature gradient. The temperature gradient first initiates variance of the internal kinetic energy in the degree of freedom corresponding to the direction of the temperature gradient. (For the case in Fig.\ref{Fig15} the temperature shows gradient in the radial direction. This gradient first initiates the variance of the mean kinetic energy $\int d\mathbf{v} f (v_r - u_r)^2 /2$.) Then, part of internal kinetic energy variance is transferred to other degrees of freedoms via collisions of molecules. Then, the internal kinetic energy in this degree of freedom further varies according to the temperature gradient, and so on. Only when the temperature gradient vanishes, the system can attain its thermodynamic equilibrium, i.e. the internal kinetic energy in different degrees of the freedom equal to each other.

\subsection{Recovering of the distribution function}

\begin{figure}[tbp]
\center\includegraphics*[bbllx=0pt,bblly=177pt,bburx=595pt,bbury=666pt,angle=0,width=0.40\textwidth]{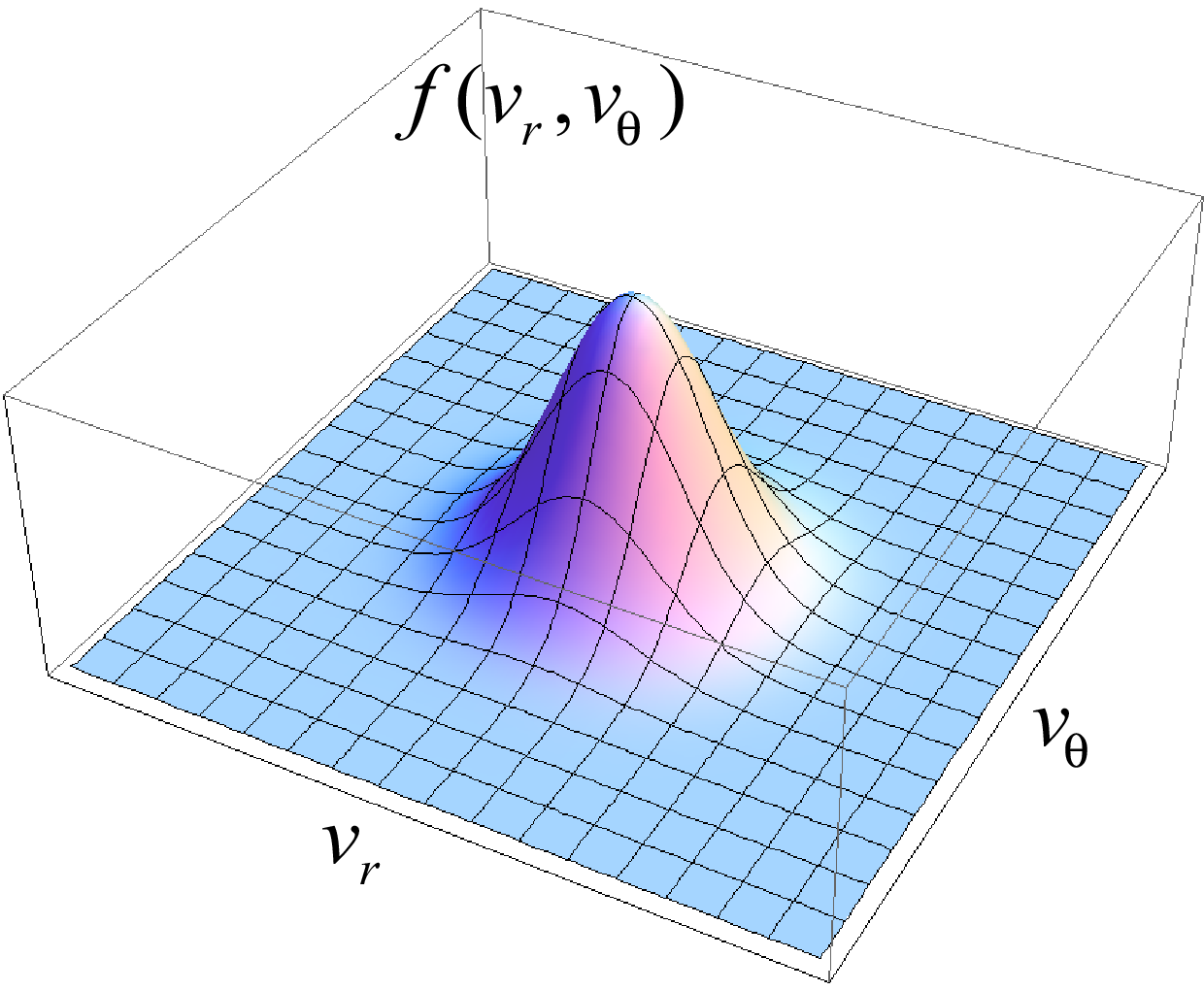}
\caption{The sketch of the Maxwellian distribution function in velocity space ($v_{r}$,$v_{\theta}$).}
\label{Fig24}
\end{figure}

When the system is in a thermodynamic equilibrium state, the distribution function of the particle velocity is a local Maxellian, i.e., a normal distribution, symmetric about the mean flow velocity $\mathbf{u}$. This property reflects profound symmetries of Newtonian mechanics, i.e. Galilean and scaling invariance, respectively. The local Maxwellian does not support any dissipative and transport mechanism, since these phenomena violate the aforementioned symmetries. Indeed, transport phenomena triggered by departures from local equilibria reflect into symmetry-breaking departures from the Maxwellian distribution. The maxwellian distribution is shown in Fig.\ref{Fig24}.
%%%%%%%%%%%%%%%%%%%%%%%%%%%%%%%%%%%%%%%%%%%%%%%%%%%%%%%%%%%%%%%%%%%%
%%%%%%%%%%%%%%%%%%%%%%%%%%%%%%%%%%%%%%%%%%%%%%%%%%%%%%%%%%%%%%%%%%%%
\begin{figure}[tbp]
\center\includegraphics*[bbllx=33pt,bblly=61pt,bburx=560pt,bbury=195pt,angle=0,width=0.99\textwidth]{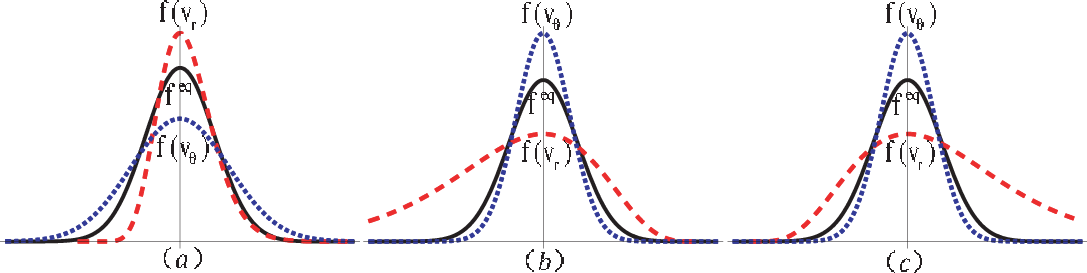}
\caption{The sketch of the Maxwellian and actual distribution functions versus velocity $v_{r}$ and $v_{\theta}$, respectively.
Figures (a)-(c) show the distribution functions at the rarefaction front, the material interface and the shock front, respectively.
The long-dashed, shot-dashed and solid lines are for distribution functions $f(v_{r})$, $f(v_{\theta })$ and $f^{eq}$, respectively.}
\label{Fig25}
\end{figure}
%%%%%%%%%%%%%%%%%%%%%%%%%%%%%%%%%%%%%%%%%%%%%%%%%%%%%%%%%%%%%%%%%%%%
\begin{figure}[tbp]
\center\includegraphics*[bbllx=0pt,bblly=0pt,bburx=595pt,bbury=200pt,angle=0,width=0.99\textwidth]{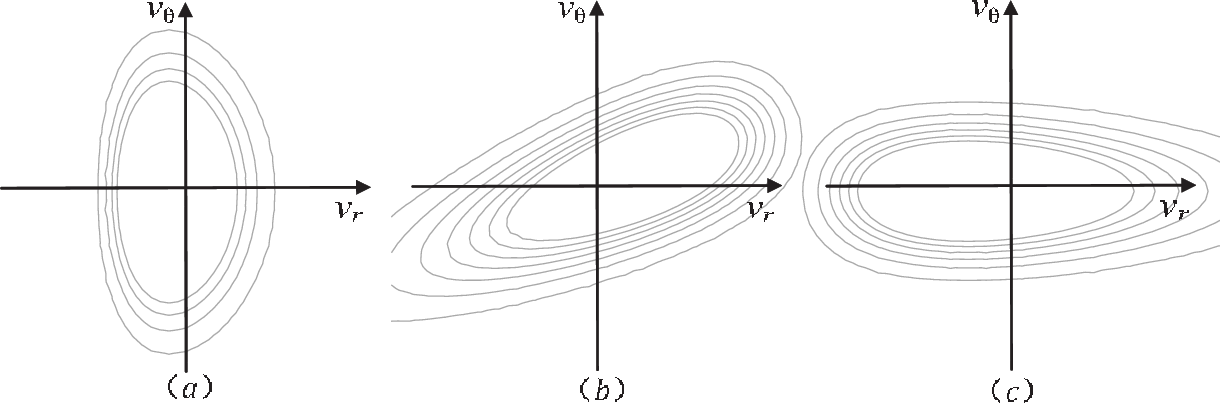}
\caption{The sketch of the contours of actual distribution functions in velocity space ($v_{r}$,$v_{\theta}$).
Figure (a)-(c) show the recovered distribution function contours at the rarefaction front, the material interface and the shock front, respectively. }
\label{Fig26}
\end{figure}
%%%%%%%%%%%%%%%%%%%%%%%%%%%%%%%%%%%%%%%%%%%%%%%%%%%%%%%%%%%%%%%%%%%%
\begin{figure}[tbp]
\center\includegraphics*[bbllx=0pt,bblly=290pt,bburx=595pt,bbury=468pt,angle=0,width=0.95\textwidth]{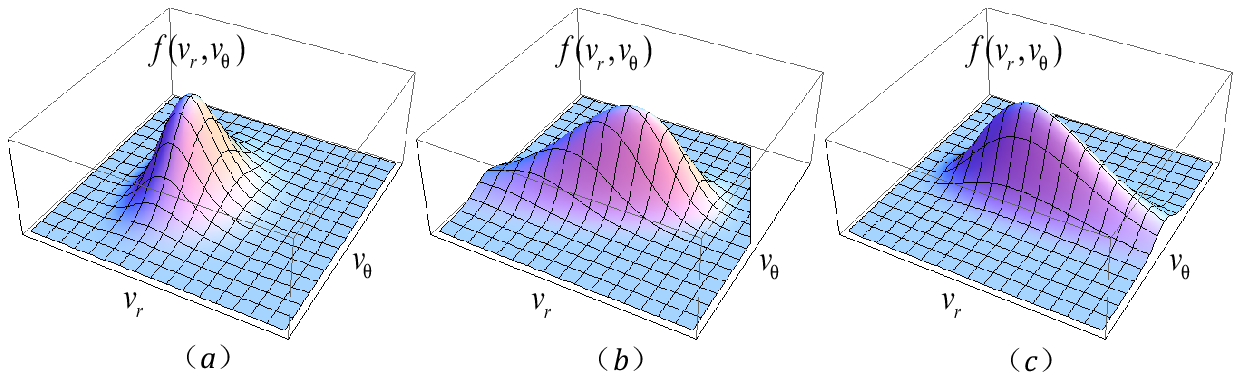}
\caption{The sketches of the actual distribution functions in velocity space ($v_{r}$,$v_{\theta}$). Figures (a)-(c) show the recovered distribution functions at the rarefaction front, the material interface and the shock front, respectively. }
\label{Fig27}
\end{figure}
%%%%%%%%%%%%%%%%%%%%%%%%%%%%%%%%%%%%%%%%%%%%%%%%%%%%%%%%%%%%%%%%%%%%

From the simulation results of the deviations $\mathbf{\Delta}^*_m$, we can draw qualitative information on the actual distribution function. As an example, we consider the above-mentioned case without initial perturbation at the material interface, and recover qualitatively the actual distribution function. The main steps are given below.

We first consider the actual functions $f(v_{r})$ and $f(v_{\theta})$ at the rarefaction front. It's easy to find in Fig.\ref{Fig19} (a) that $\mathbf{\Delta}^*_{2,rr}$ shows a negative peak and $\mathbf{\Delta}^*_{2,\theta \theta}$ shows a positive peak with the
same amplitude. Up to this step, we can imagine that the distribution function $f(v_{r})$ is \textquotedblleft thinner\textquotedblright and $f(v_{\theta })$ is \textquotedblleft fatter\textquotedblright than the Maxwellian. The peak of $f(v_{r})$ is higher and the peak of $f(v_{\theta})$ is lower than that of the Maxwellian. $\mathbf{\Delta}^*_{4,2}$ in Fig.\ref{Fig19} (d) shows complementary information to $\mathbf{\Delta}^*_{2}$ in Fig.\ref{Fig19} (a). According to $\mathbf{\Delta}^*_{3}$ in Fig.\ref{Fig19} (b) and $\mathbf{\Delta}^*_{3,1}$ in Fig.\ref{Fig19} (c), we can obtain that $f(v_{\theta })$ is symmetric, while the $f(v_{r})$ is asymmetric. The portion for $v_{r}>0$ is \textquotedblleft fatter\textquotedblright than that for $v_{r}<0$. This is often called ``positive skewness". Figure \ref{Fig25} (a) shows a sketch of the actual distribution functions $f(v_{r})$, $f(v_{\theta})$ and the Maxwellian $f^{eq}$. Here $f^{eq}=f^{eq}(v_{r})=f^{eq}(v_{\theta })$ due to the symmetry of the Maxwellian. A sketch of the distribution functions around the shock wave is shown in Fig.\ref{Fig25} (c), where $f(v_{r})$ is \textquotedblleft fatter\textquotedblright and $f(v_{\theta})$ is \textquotedblleft thinner\textquotedblright than the Maxwellian. The peak of $f(v_{r})$ is lower and the peak of $f(v_{\theta})$ is higher than that of the Maxwellian. And $f(v_{\theta})$ is symmetric while $f(v_{r})$ is asymmetric. The portion for $v_{r}>0$ is \textquotedblleft fatter\textquotedblright and the portion for $v_{r}<0$ is \textquotedblleft thinner\textquotedblright. Similarly, a sketch of the actual distribution functions at the materia interface is shown in Fig.\ref{Fig25} (b).

Secondly, we study the contours of the actual distribution function in two-dimensional velocity space ($v_{r}$,$v_{\theta}$). It's clear that the values of $\mathbf{\Delta}^*_{2,r \theta}$ in Fig.\ref{Fig19} (a) and Fig.\ref{Fig19} (i) equal to zero, which implies that the contours of the actual distribution function at the rarefaction and the shock waves ought to be symmetric about $v_{r}$ axis or/and $v_{\theta}$ axis. With this mind that $f(v_{\theta })$ is symmetric at the two interfaces, we can confirm that $v_{r}$ axis is the symmetric axis of the two contours. Figures \ref{Fig19} (d) and (l) show consistent information. $\mathbf{\Delta}^*_{2,r \theta}$ in Fig.\ref{Fig19} (e) shows a positive peak, which implies that, at the material interface, the contour is not symmetric about the $v_{r}$ or $v_{\theta}$ axis. Because the shear viscous effects are pronounced, the actual distribution function is relatively complex. Figure \ref{Fig26} shows, from left to right, the sketches of contours of the actual distribution function at the interfaces of rarefaction, material and shock.

Finally, by combining the results of the above two steps, we obtain the qualitative curves for the actual distribution functions at the three interfaces. The sketches are shown in Fig.\ref{Fig27}. Figures (a)-(c) are for the rarefaction front, the material interface and the shock front, respectively. It should be pointed out that, since only $7$ moment relations are used in the current LB model, only part of the information on the actual distribution function can be qualitatively recovered.

\section{Conclusions and discussions}

A polar coordinate lattice Boltzmann kinetic model for compressible flows is presented. A combined scheme is proposed for solving the LB equation. The convection term is solved via a modified Warming-Beam scheme where a switch function is introduced. The temporal evolution is calculated analytically. The new model works for both subsonic and supersonic flows. Consequently, it can be used to study complex flows under strong impact or shock. The new model is validated and verified via typical benchmark tests, (i) the rotational flow, (ii) the Kelvin-Helmholtz instability, (iii) the stable shock tube problem, and (iv) the Richtmyer-Meshkov instability. Among them the latter two can not be simulated by the previous PCLB model\cite{Watari2011}. Even for the former two cases where the previous model \cite{Watari2011} works, the simulation results by the new model appear to be more accurate.

Choosing computational domain and designing boundary conditions play an important role in numerical experiments. For annular systems showing periodic behaviors in the circumferential direction, one can pick out only one period of the domain for simulations. In such a case, the two boundaries in the circumferential direction are treated with periodic conditions. The two boundaries in the radial direction should be treated carefully according to the specific situation under investigation. The simplest microscopic radial boundary conditions assume that the system at the inner and outer boundaries are in thermodynamic equilibriums. The more accurate microscopic radial boundary conditions take also into account the deviation from thermodynamic equilibrium. The deviation from thermodynamic equilibrium can be obtained via extrapolation scheme from values at the neighboring lattice nodes inside the system \cite{Guo2002}.

Compared with the continuum based model for compressible flow, such as the Navier-Stokes equations, the LB kinetic model presented in this work can be used to investigate a rich variety of non-equilibrium effects of the system due to its deviations from thermodynamic equilibrium. Both the current LB model and the Gas Kinetic Scheme(GKS) presented in Ref.\cite{XuKun2000,XuKun2003} are relevant to the Boltzmann equation. But they are significantly different. In the current LB kinetic model, the distribution function contains all the physical information. It describes the equilibrium and nonequilibrium phenomena of the system. One can observe the nonequilibrium effects by inspecting the high-order moments of distribution function. The LB code describes the evolution of the discrete distribution function. The GKS is a kind of finite volume scheme where the fluxes are evaluated from the distribution function.

To show the merit of LB kinetic model over the traditional methods based on continuum assumption, we studied the macroscopic behaviors of the system due to deviating from thermodynamic equilibrium around three kinds of interfaces, the shock wave, the rarefaction wave and the material interface, for two specific cases. In one of the two cases, the material interface is initially perturbed and consequently the RM instability occurs. It is found that, the macroscopic effects of deviating from thermodynamic equilibrium around the material interface are greatly different from those around the mechanical interfaces. The initial perturbation at the material interface results in more pronounced two-dimensional effects and enhanced coupling of molecular motions in different degrees of freedom. The system deviates much more from thermodynamic equilibrium around the shock wave than around the material interface and the rarefaction wave. By comparing each component of the high-order moments and its value in equilibrium, we can draw qualitatively the main information of the actual distribution function which determines the macroscopic behaviors. These results deepen our understanding on the mechanical and material interfaces from a more fundamental level, and present valuable information for improving the macroscopic modeling. More systematic study on the non-equilibrium effects in RM and KH instabilities is in progress.

%%%%%%%%%%%%%%%%%%%%%%%%%%%%%%%%%%%%%%%%%%%%%%%%%%%%%%%%%%%%%%%%%%%%

\section*{Acknowledgements}

The authors thank Prof. Guoxi Ni for many helpful discussions. AX and GZ acknowledge support of the Science Foundations of CAEP [under Grant Nos. 2012B0101014 and 2011A0201002] and the Foundation of State Key Laboratory of
Explosion Science and Technology [under Grant No. KFJJ14-1M]. AX, GZ, YL and CL acknowledge support of National Natural Science Foundation of China [under Grant Nos.11074300, 11202003, and 91130020]. YL and CL acknowledge support of National Basic Research Program of China (Grant No. 2013CBA01504).

\section*{Appendix}

The LB equation \eqref{combined_scheme} can be written as below,
\begin{equation}
f_{ki}^{t+\Delta t}=term^{t}+term^{r}+term^{\theta }  \label{combined_term}
\end{equation}
with
\begin{equation}
\left\{
\begin{array}{ccl}
term^{t} & = & \exp (-\frac{\Delta t}{\tau })[f_{ki}^{t}-f_{ki}^{eq}+f_{ki}^{eq}\exp (\frac{\Delta t}{\tau })] \\
term^{r} & = & -[C_{r}+\frac{1}{2}C_{r}(1-C_{r})(1-S(\eta _{r}))(1-\eta _{r})]\delta _{r} \\
term^{\theta} & = & -[C_{\theta}+\frac{1}{2}C_{\theta }(1-C_{\theta})(1-S(\eta_{\theta}))(1-\eta _{\theta })]\delta_{\theta} \text{.}
\end{array}
\right.   \notag
\end{equation}
Using Taylor expansion for the two sides of Eq.\eqref{combined_term} in the case $C_{r}>0$, $C_{\theta }>0$, $\eta _{r}>0$ and $\eta _{\theta }>0$, we get
\begin{equation}
f_{ki}^{t+\Delta t}=f_{ki}+\frac{\partial f_{ki}}{\partial t}\Delta t+\frac{1}{2}\frac{\partial ^{2}f_{ki}}{\partial t^{2}}\Delta t^{2}+O(\Delta t^{3})\text{,}  \label{leftside}
\end{equation}
\begin{equation}
term^{t}=f_{ki}^{eq}+(f_{ki}^{t}-f_{ki}^{eq})[1+(-\frac{\Delta t}{\tau })+\frac{1}{2}(-\frac{\Delta t}{\tau })^{2}+O(\Delta t^{3})]\text{,}
\label{rightside1}
\end{equation}
\begin{equation}
\begin{array}{ccl}
term^{r} & = &
-C_{r}(f_{ki,ir}^{t}-f_{ki,ir-1}^{t})-C_{r}(1-C_{r})(f_{ki,ir}^{t}-2f_{ki,ir-1}^{t}%
+f_{ki,ir-2}^{t})g(\eta _{r}) \\
& = & -\upsilon _{kir}\frac{\partial f_{ki}}{\partial r}\Delta t%
+\frac{1}{2}{\frac{\partial ^{2}f_{ki}}{\partial r^{2}}[1-g(\eta _{r})]%
{\upsilon_{kir}}\Delta r\Delta t}
+\frac{{{\upsilon _{kir}^{2}}}}{2}{\frac{\partial ^{2}f_{ki}}{\partial r^{2}}g(\eta _{r})%
\Delta t^{2}}+\Delta tO(\Delta r^{2}) \text{,}
\end{array}
\label{rightside2}
\end{equation}
\begin{equation}
\begin{array}{ccl}
term^{\theta } & = & -C_{\theta }(f_{ki,i\theta }^{t}-f_{ki,i\theta-1}^{t})
-C_{\theta }(1-C_{\theta })(f_{ki,i\theta }^{t}
-2f_{ki,i\theta-1}^{t}+f_{ki,i\theta -2}^{t})g(\eta _{\theta }) \\
& = & -\frac{\upsilon _{ki\theta }}{r}\frac{\partial f_{ki}}{\partial \theta
}\Delta t+\frac{\upsilon _{ki\theta }}{2r}{\frac{\partial ^{2}f_{ki}}{\partial \theta ^{2}}}[{1-g(\eta _{\theta })}]{\Delta \theta \Delta t}
+{{\frac{{{\upsilon _{ki\theta }^{2}}}}{2r^{2}}}\frac{\partial ^{2}f_{ki}}{\partial \theta ^{2}}g(\eta _{\theta })\Delta t^{2}}+\Delta tO(\Delta \theta^{2}) \text{,}
\end{array}
\label{rightside3}
\end{equation}
with
\begin{equation}
g(\eta_{r})=1-S(\eta_{r})
=2\frac{f_{ki,ir}-f_{ki,ir-1}}{f_{ki,ir}-f_{ki,ir-2}}
=\frac{\frac{\partial f_{ki}}{\partial r}-\frac{1}{2}%
\frac{\partial ^{2}f_{ki}}{\partial r^{2}}\Delta r+\frac{1}{6}\frac{\partial
^{3}f_{ki}}{\partial r^{3}}\Delta r^{2}+O(\Delta r^{3})}{\frac{\partial
f_{ki}}{\partial r}-\frac{\partial ^{2}f_{ki}}{\partial r^{2}}\Delta r+\frac{%
2}{3}\frac{\partial ^{3}f_{ki}}{\partial r^{3}}\Delta r^{2}+O(\Delta r^{3})}
\label{g_eta_r}
\text{,}
\end{equation}
\begin{equation}
g(\eta_{\theta })=1-S(\eta_{\theta })
=2\frac{f_{ki,i\theta}-f_{ki,i\theta -1}}{f_{ki,i\theta }-f_{ki,i\theta -2}}
=\frac{\frac{\partial f_{ki}}{\partial \theta }-\frac{1}{2}\frac{\partial ^{2}f_{ki}}{\partial
\theta ^{2}}\Delta \theta +\frac{1}{6}\frac{\partial ^{3}f_{ki}}{\partial
\theta ^{3}}\Delta \theta ^{2}+O(\Delta \theta ^{3})}{\frac{\partial f_{ki}}{%
\partial \theta }-\frac{\partial ^{2}f_{ki}}{\partial \theta ^{2}}\Delta
\theta +\frac{2}{3}\frac{\partial ^{3}f_{ki}}{\partial \theta ^{3}}\Delta
\theta ^{2}+O(\Delta \theta ^{3})}
\label{g_eta_theta}
\text{.}
\end{equation}
Via Taylor expansion, Eqs.\eqref{g_eta_r}-\eqref{g_eta_theta} give
\begin{equation}
g(\eta_{r})=1+O(\Delta r)
\label{g_Talor_r}
\text{,}
\end{equation}
\begin{equation}
g(\eta_{\theta})=1+O(\Delta \theta)
\label{g_Talor_theta}
\text{.}
\end{equation}
Substituting Eqs.\eqref{leftside}-\eqref{rightside3} into \eqref{combined_term}, we get
\begin{equation}
\begin{array}{l}
\frac{\partial f_{ki}}{\partial t}+v_{kir}\frac{\partial f_{ki}}{\partial r}
+\frac{1}{r}v_{ki\theta }\frac{\partial f_{ki}}{\partial \theta }
=-\frac{1}{\tau }[f_{ki}-f_{ki}^{eq}] \\
-\frac{1}{2}\frac{\partial ^{2}f_{ki}}{\partial t^{2}}\Delta t+\frac{1}{%
2\tau ^{2}}(f_{ki}^{t}-f_{ki}^{eq})\Delta t+\frac{1}{2}\upsilon _{kir}^{2}%
\frac{\partial ^{2}f_{ki}^{t}}{\partial r^{2}}{g(\eta _{r})}\Delta t+\frac{%
\upsilon _{ki\theta }^{2}}{2r^{2}}\frac{\partial ^{2}f_{ki}}{\partial \theta
^{2}}{g(\eta _{r})}\Delta t \\
+\frac{\upsilon_{kir}}{2}\frac{\partial ^{2}f_{ki}}{\partial r^{2}}[1-{g(\eta _{r})}]\Delta r
+\frac{\upsilon _{ki\theta }}{2r}\frac{\partial^{2}f_{ki}}{\partial \theta ^{2}}%
[1-{g(\eta _{\theta })}]\Delta \theta  \\
+O(\Delta t^{2})+O(\Delta r^{2})+O(\Delta \theta ^{2})
\text{.}
\end{array}
\label{PCLBE_MWB}
\end{equation}
Substituting Eqs.\eqref{g_Talor_r}-\eqref{g_Talor_theta} into \eqref{PCLBE_MWB}, we get
\begin{equation}
\begin{array}{l}
\frac{\partial f_{ki}}{\partial t}+v_{kir}\frac{\partial f_{ki}}{\partial r}
+\frac{1}{r}v_{ki\theta }\frac{\partial f_{ki}}{\partial \theta }
=-\frac{1}{\tau }[f_{ki}-f_{ki}^{eq}] \\
-\frac{1}{2}\frac{\partial ^{2}f_{ki}}{\partial t^{2}}\Delta t+\frac{1}{%
2\tau ^{2}}(f_{ki}^{t}-f_{ki}^{eq})\Delta t+\frac{1}{2}\upsilon _{kir}^{2}%
\frac{\partial ^{2}f_{ki}^{t}}{\partial r^{2}}\Delta t+\frac{%
\upsilon _{ki\theta }^{2}}{2r^{2}}\frac{\partial ^{2}f_{ki}}{\partial \theta
^{2}}\Delta t \\
+O(\Delta t^{2})+O(\Delta r^{2})+O(\Delta \theta ^{2})
\text{.}
\end{array}
\label{PCLB_error}
\end{equation}
Comparing with Eq.\eqref{PolarLBE1}, the above equation has a first-order truncation error in the case $C_{r}>0$, $C_{\theta }>0$, $\eta _{r}>0$ and $\eta _{\theta }>0$. This conclusion is also suitable for other cases. Consequently, our combined scheme has first-order accuracy as a whole.

Via the Chapman-Enskog expansion, it's found that the LB equation \eqref{PCLB_error} presents the following equations
\begin{equation}
\begin{array}{l}
\frac{\partial \rho }{\partial t}+\nabla \cdot (\rho \mathbf{u})=-\frac{\Delta t}{2}\frac{\partial ^2\rho }{\partial t^2}+\frac{\Delta t}{2}\frac{\partial^2}{\partial r^2}(\rho E+\rho u_r^2) \\
+\frac{\Delta t}{2r^2}[\frac{\partial ^2(\rho E+\rho u_{\theta }^2)}{\partial \theta ^2}
+4\varepsilon \frac{\partial \rho u_ru_{\theta }}{\partial \theta }
+2\rho \varepsilon ^2(u_r^2-u_{\theta }^2)]
\end{array}
\label{NS1_error}
\end{equation}
\begin{equation}
\begin{array}{l}
\frac{\partial (\rho \mathbf{u})}{\partial t}+\nabla \cdot (P\mathbf{I}+\rho
\mathbf{uu})+\nabla \cdot \lbrack \mu (\nabla \cdot \mathbf{u})\mathbf{I}%
-\mu (\nabla \mathbf{u})^{T}-\mu \nabla \mathbf{u}] \\
=-\frac{\Delta t}{2\tau }\nabla \cdot \lbrack \mu (\nabla \cdot \mathbf{u})%
\mathbf{I}-\mu (\nabla \mathbf{u})^{T}-\mu \nabla \mathbf{u}] \\
-\frac{\Delta t}{2}\frac{\partial ^{2}}{\partial t^{2}}(\rho \mathbf{u})+%
\frac{\Delta t}{2}\frac{\partial ^{2}(\rho u_{r}^{3}+3\rho Eu_{r})}{\partial
r^{2}}\mathbf{e}_{r}+\frac{\Delta t}{2}\frac{\partial ^{2}(\rho u_{r}^{2}u_{\theta }+
\rho Eu_{\theta })}{\partial r^{2}}\mathbf{e}_{\theta}\\
+\frac{\Delta t}{2r^{2}}[\frac{\partial ^{2}(\rho u_{r}u_{\theta }^{2}%
+\rho Eu_{r})}{\partial ^{2}\theta }-2\frac{\partial (\rho u_{\theta }^{3}%
+3\rho Eu_{\theta })}{\partial \theta }-(\rho u_{r}u_{\theta }^{2}+\rho Eu_{r})]%
\mathbf{e}_{r} \\
+\frac{\Delta t}{2r^{2}}[\frac{\partial ^{2}(\rho u_{\theta }^{3}%
+3\rho Eu_{\theta })}{\partial ^{2}\theta }+2\frac{\partial (\rho u_{r}u_{\theta}^{2}%
+\rho Eu_{r})}{\partial \theta }-(\rho u_{\theta }^{3}%
+3\rho Eu_{\theta})]\mathbf{e}_{\theta } \\
+\frac{2\Delta t\varepsilon }{r^{2}}[\frac{\partial (\rho u_{r}^{2}u_{\theta}%
+\rho Eu_{\theta })}{\partial \theta }-(\rho u_{r}u_{\theta }^{2}%
+\rho Eu_{r})]\mathbf{e}_{r} \\
+\frac{2\Delta t\varepsilon }{r^{2}}[\frac{\partial (\rho u_{r}u_{\theta}^{2}%
+\rho Eu_{r})}{\partial \theta }+(\rho u_{r}^{2}u_{\theta }%
+\rho Eu_{\theta })]\mathbf{e}_{\theta } \\
+\frac{\Delta t\varepsilon ^{2}}{r^{2}}[(\rho u_{r}^{3}+3\rho Eu_{r})\mathbf{e}_{r}%
+(\rho u_{r}^{2}u_{\theta }+\rho Eu_{\theta })\mathbf{e}_{\theta }] \\
-\frac{\Delta t\varepsilon ^{2}}{r^{2}}[(\rho u_{r}u_{\theta }^{2}%
+\rho Eu_{r})\mathbf{e}_{r}+(\rho u_{\theta }^{3}+3\rho Eu_{\theta })\mathbf{e}_{\theta }]
\end{array}
\label{NS2_error}
\end{equation}
\begin{equation}
\begin{array}{l}
\frac{\partial }{\partial t}(\rho E+\frac{1}{2}\rho u^{2})+\nabla \cdot
\lbrack \rho \mathbf{u}(E+\frac{1}{2}u^{2}+\frac{P}{\rho })] \\
-\nabla \cdot \lbrack \kappa ^{^{\prime }}\nabla E+\mu \mathbf{u}\cdot
(\nabla \mathbf{u})-\mu \mathbf{u}(\nabla \cdot \mathbf{u})+\frac{1}{2}\mu
\nabla u^{2}] \\
=\frac{\Delta t}{2\tau }\nabla \cdot \lbrack \kappa ^{^{\prime }}\nabla E%
+\mu \mathbf{u}\cdot (\nabla \mathbf{u})%
-\mu \mathbf{u}(\nabla \cdot \mathbf{u})+\frac{1}{2}\mu \nabla u^{2}] \\
-\frac{\Delta t}{2}\frac{\partial ^{2}}{\partial t^{2}}(\rho E+\frac{1}{2}%
\rho u^{2})+\frac{\Delta t}{2}\frac{\partial ^{2}}{\partial r^{2}}%
[\rho E(2E+\frac{u^{2}}{2})+\rho u_{r}^{2}(3E+\frac{u^{2}}{2})] \\
+\frac{\Delta t}{2r^{2}}\{\frac{\partial ^{2}}{\partial \theta ^{2}}%
[\rho E(2E+\frac{u^{2}}{2})+\rho u_{\theta }^{2}(3E+\frac{u^{2}}{2})] \\
+4\frac{\partial }{\partial \theta }[\rho u_{r}u_{\theta }(3E+\frac{u^{2}}{2})]%
+2\rho (3E+\frac{u^{2}}{2})(u_{r}^{2}-u_{\theta }^{2})\}\text{.}
\end{array}
\label{NS3_error}
\end{equation}
Comparing the above three equations with Navier-Stokes equations in Eqs.\eqref{NS_1}-\eqref{NS_3}, it's easy to get the numerical errors in the right sides of Eqs.\eqref{NS1_error}-\eqref{NS3_error}. It is clear that the numerical errors reduce with decreasing $\Delta t$. Consequently, the Galilean invariance problem vanishes when $\Delta t$ approaches zero.

There are two kinds of discretizations in the current LB model. One kind is for the temporal and spatial derivatives which brings the truncation errors as mentioned above. The other kind is for the velocity space. The present model is a kind of FDLB model, which is quite different from the standard LB model where the discretization of the velocity space is combined with the discretizations of the space and time\cite{Nie2008}. It is also meaningful to mention that, when shocks exist in the compressible flow system, they proceed much faster than effects resulting from violations of Galilean invariance. In other words, the shocking effects play a dominant role in the concerned time scale, and the Galilean invariance problem can be negligible.


\begin{thebibliography}{99}

\bibitem{Succi-Book} S. Succi, \textit{The Lattice Boltzmann Equation for Fluid Dynamics and Beyond}, Oxford University Press, New York, (2001).

\bibitem{Alexander1992} F. J. Alexander, H. Chen, S. Chen and G. D. Doolen, Phys. Rev. A 46, 1967 (1992).

\bibitem{Yan1999} G. Yan, Y. Chen, S. Hu, Phys. Rev. E 59, 454 (1999).

\bibitem{Sun1998} C. H. Sun, Phys. Rev. E 58, 7283 (1998).

\bibitem{Sun2003} C. Sun and A. T. Hsu, Phys. Rev. E 68, 016303 (2003).

\bibitem{Cao1997} N. Cao, S. Chen, S. Jin, and D. Martinez, Phys. Rev. E 55, R21 (1997).

\bibitem{Kataoka2004a} T. Kataoka and M. Tsutahara, Phys. Rev. E 69, 056702 (2004).

\bibitem{Kataoka2004b} T. Kataoka and M. Tsutahara, Phys. Rev. E 69, 035701(R)(2004).

\bibitem{Watari2003} M. Watari and M. Tsutahara, Phys. Rev. E 67, 036306 (2003).

\bibitem{Watari2004} M. Watari and M. Tsutahara, Phys. Rev. E 70, 016703 (2004).

\bibitem{Watari2007} M. Watari, Physica A 382, 502 (2007).

\bibitem{Xu2005PRE} A. Xu, Phys. Rev. E 71, 066706 (2005).

\bibitem{Xu2005EPL} A. Xu, Europhys. Lett. 69, 214 (2005).

\bibitem{XGL1} A. Xu, G. Gonnella, and A. Lamura, Phys. Rev. E 67, 056105 (2003).

\bibitem{XGL2} A. Xu, G. Gonnella, A. Lamura, G. Amati, and F. Massaioli, Europhys. Lett. 71, 651 (2005).

\bibitem{XGL3} A. Xu, G. Gonnella, and A. Lamura, Phys. Rev. E 74, 011505 (2006).

\bibitem{LB2MPhase2011PRE} Y. Gan, A. Xu, G. Zhang, Y. Li and H. Li, Phys. Rev. E 84, 046715 (2011).

\bibitem{LB2MPhase2012EPL} Y. Gan, A. Xu, G. Zhang, and Y. Li, Europhys. Lett. 97, 44002 (2012).

\bibitem{LB2MPhase2012FrontPhys} Y. Gan, A. Xu, G. Zhang, and Y. Li, Front. Phys. 7(4), 481 (2012)

\bibitem{XuPan2007} X. Pan, A. Xu, G. Zhang, and S. Jiang, Int. J. Mod. Phys. C 18, 1747 (2007).

\bibitem{XuGan2008CTP} Y. Gan, A. Xu, G. Zhang, and Y. Li, Commun. Theor. Phys. 50, 201 (2008).

\bibitem{XuGan2008PhysA} Y. Gan, A. Xu, G. Zhang, X. Yu, and Y. Li, Physica A 387, 1721 (2008).

\bibitem{XuGan2011CTP} Y. Gan, A. Xu, G. Zhang, and Y. Li, Commun. Theor. Phys. 56, 490 (2011).

\bibitem{LB2KHI2011} Y. Gan, A. Xu, G. Zhang, and Y. Li, Phys. Rev. E 83, 056704 (2011).

\bibitem{XuChen2009} F. Chen, A. Xu, G. Zhang, Y. Gan, C. Tao, and Y. Li, Commun. Theor. Phys. 52, 681 (2009).

\bibitem{MRT2010EPL} F. Chen, A. Xu, G.Zhang, Y. Li, S. Succi, EuroPhys. Lett. 90, 54003 (2010).

\bibitem{XuChen2010} F. Chen, A. Xu, G. Zhang, Y. Li, Commun. Theor. Phys. 54, 1121, (2010).

\bibitem{MRT2011CTP} F. Chen, A. Xu, G.Zhang, Y. Li, Commun. Theor. Phys. 55, 325 (2011).

\bibitem{MRT2011PLA} F. Chen, A. Xu, G.Zhang, Y. Li, Phys. Lett. A 375, 2129 (2011).

\bibitem{XuChen2011} F. Chen, A. Xu, G. Zhang, Y. Li, Commun. Theor. Phys. 56, 333, (2011).

\bibitem{MRT2011TAML} F. Chen, A. Xu, G.Zhang, Y. Li, Theroe. \& Appl. Mech. Lett. 1, 052004 (2011).

\bibitem{Review2012} A. Xu, G. Zhang, Y. Gan, F. Chen, and X. Yu, Front. Phys. 7(5), 582 (2012)

\bibitem{Nannelli1992} F. Nannelli, S. Succi, J. Stat. Phys. 68, 401 (1992).

\bibitem{Succi1995} S. Succi, G. Amati, and R. Benzi, J. Stat. Phys. 81, 5 (1995).

\bibitem{Amati1997} G. Amati, S. Succi and R. Benzi, Fluid Dyn. Res. 19, 289 (1997).

\bibitem{Peng1998} G. Peng, H. Xi, C. Duncan and SH. Chou, Phys. Rev. E 58, R4125 (1998).

\bibitem{Peng1999} G. Peng, H. Xi, C. Duncan and SH. Chou, Phys. Rev. E 59, 4676 (1999)

\bibitem{Ubertini2003} S. Ubertini, G. Bella and S. Succi, Phys. Rev. E 68, 016701 (2003)

\bibitem{He1997} X. He, G. Doolen, J. Comput. Phys. 134, 306 (1997).

\bibitem{Mei1998} R. Mei, W. Shyy, J. Comput. Phys. 143, 426 (1998).

\bibitem{Halliday2001} I. Halliday, L. A. Hammond, C. M. Care, K. Good, and A. Stevens, Phys. Rev. E 64, 011208 (2001).

\bibitem{Premnath2005} K. N. Premnath and J. Abraham, Phys. Rev. E 71, 056706 (2005).

\bibitem{Asinari2010} P. Asinari, S. C. Mishra and R. Borchiellini, Numerical Heat Transfer B 57, 126 (2010).

\bibitem{Watari2011} M. Watari, Commun. Comput. Phys. 9, 1293 (2011).

\bibitem{bgk_1954} P. Bhatnagar, E. P. Gross, and M. K. Krook, Phys. Rev. 94, 511 (1954).

\bibitem{Guo2002} Z. Guo, C. Zheng, B. Shi, Phys. Fluids, 14, 2007 (2002).

\bibitem{Richtmyer1960} R. D. Richtmyer, Comm. Pure Appl. Math. 13, 297 (1960).

\bibitem{Meshkov69} E. E. Meshkov, Sov. Fluid Dyn. 4, 101 (1969).

\bibitem{Benjamin92} R. F. Benjamin, \textit{Advances in Compressible Turbulent Mixing}, edited by W. P. Dannevik, A. C. Buckingham, and C. E. Leith (USGPO, Washington, 1992).

\bibitem{Zhang1997} Q. Zhang, S. Sohn, Appl. Math. Lett., 10, 121 (1997).

\bibitem{RMI_Review2002} M. Brouillette, Annu. Rev. Fluid Mech. 34, 445 (2002).

\bibitem{Dellar2013} P. Dellar, Comput. Math. Applic. 65, 129 (2013).

\bibitem{XuKun2000} Y. Lian, K. Xu, J. Comput. Phys., 163(2), 349, (2000)

\bibitem{XuKun2003} K. Xu, X. He, J. Comput. Phys., 190(1), 100, (2003)

\bibitem{Nie2008} X. B. Nie, X. Shan, and H. Chen, Europhys. Lett. 81, 34005 (2008).

\end{thebibliography}
\end{document}